\documentclass[twocolumn]{aa}  

\usepackage{natbib}
\bibpunct{(}{)}{;}{a}{}{,}

\usepackage{graphicx}
\usepackage{txfonts}
\usepackage{lipsum}
\usepackage{subcaption}         
\usepackage{lscape}             
\usepackage{placeins}           

\usepackage{mathtools}
\usepackage{multirow}

\begin{document}

   \title{Fizzy water ice in space: CO$_2$ adsorption, binding energies and its fate in a protoplanetary disk}

   \subtitle{}

\author{A. Bulik\inst{1}\fnmsep $^{\star}$
        \and V. Bariosco\inst{1,2} 
        \and E. Mates-Torres\inst{1} 
        \and P. Ugliengo\inst{2}
        \and K. Furuya\inst{3} 
        \and C. Ceccarelli\inst{4} 
        \and A. Rimola\inst{1,5}\fnmsep $^{\star}$
        }

\institute{Departament de Química, Universitat Autònoma de Barcelona, 08193 Bellaterra, Catalonia, Spain
            \and Dipartimento di Chimica and NIS – Nanostructured Interfaces and Surfaces – Centre, 
            Università degli Studi di Torino, via P. Giuria 7, 10125, Torino, Italy
            \and RIKEN Pioneering Research Institute, 2-1 Hirosawa, Wako-shi, Saitama 351-0198, Japan
            \and Univ. Grenoble Alpes, IPAG, F-38000 Grenoble, France
            \and Accademia delle Scienze di Torino, Via Maria Vittoria, 3, 10123 Torino, Italy
           }

\date{Received September 30, 20XX}

 
  \abstract
   {CO$_2$ is the third most abundant ice component found on dust grains in star-forming regions and a common ingredient of exoplanet atmospheres. Characterization of its adsorption properties on ices through the binding energy (BE) is essential for accurate astrochemical modelling and understanding chemical inheritance in planet formation.}
   {We aim to derive an accurate and statistically relevant BE distribution of CO$_2$ on water ices, considering both single molecule and multi-molecule adsorption scenarios. 
   Our goal is to understand the impact of the BE distribution on the abundance of gaseous and frozen CO$_2$ in a generic protoplanetary disk and the spectral absorption features of frozen CO$_2$.}
   {The ACO-FROST procedure is used for computing the BE distribution, where CO$_2$ molecules are adsorbed on several sites of an amorphous water ice grain model.
   The BEs are computed using an ONIOM scheme: ON(DLPNO-CCSD(T)/aug-cc-pvtz:GFN2)//ON(B97-3c:GFN2), which was chosen based on a detailed benchmark.}
   {The BEs of CO$_2$ follow a bimodal Gaussian distribution characterised by the following parameters: $\mu_1$ = 1648K, $\sigma_1$=229K, $\mu_2$=2339K, $\sigma_2$=274K.
   For each BE bin, the pre-exponential factor was estimated using two models (Tait and Campbell) and the Polanyi-Wigner relationship. 
   Comparison with previous studies, both experimental and computational, show good agreement on the range of the BEs.
   The impact of the adsorption on water ice on the spectral features of CO$_2$ molecule is evaluated.
   The coverage simulation shows the non-wetting properties of CO$_2$ on the water ice surface.
   We finally discuss the impact of using a BE distribution and different pre-exponential factors to calculate the partitioning between the ice and gas in a generic protoplanetary disk. 
   }
   {We confirm that the use of BE distribution to model the gas and ice fractionation in a protoplanetary disk causes the gas fraction to be significantly more extended.
   Furthermore, we show that the prefactor has a significant impact on where the snowline forms and on the final extent of the gas fraction in the disk. }


   \maketitle
   \nolinenumbers

\section{Introduction}

Carbon dioxide (CO$_2$) is the third most abundant species found on iced dust grains in star forming regions, first detected in the ice phase by \cite{dhendecourt1989}. 
Interestingly, it is often found in polar (H$_2$O-dominated) ices and less commonly in apolar (CO-dominated) ices \citep{Pontoppidan_2008,Smith2025}.
That is why the formation of CO$_2$ ice is usually assumed to be linked with its presence on water ice, the postulated mechanisms following solid-phase formation pathways such as CO + O $\rightarrow$ CO$_2$ \citep{Garrod2011,Minissale2013}, H$_2$CO + O $\rightarrow$ CO$_2$ + 2H \citep{Minissale2015}, and CO + OH $\rightarrow$ CO$_2$ + H \citep{Garrod2011,Arasa2013}. However, recent studies for this later reaction (attributed to be a major CO$_2$ formation channel) point out low efficiencies \citep{ Molpeceres2023,Ishibashi2024}, such that formation of interstellar CO$_2$ is still unclear. 

CO$_2$ has been observed in ices in various sources such as Class 0 star formation early stages and comets \citep{Brunken2024a,McClure2023, Boogert2015, Boogert_2022, Ootsubo_2012}, molecular clouds and protostellar envelopes \citep{Gerakines_1999,An_2011}, class II protoplanetary disks \citep{Sturm2023} and Galactic centre sources \citep{deGraauw1996}. 
It has also been widely detected in the gas phase \citep{Pontoppidan_2010}, around T Tauri type stars \citep{Grant_2023, Colmenares_2024, Vlasblom2025} and surrounding Herbig Ae type stars \citep{Kaeufer2025, Frediani2025}, in regions where frozen CO$_2$ sublimates.
CO$_2$ is also present in many exoplanet atmospheres \citep{Madhusudhan2019, Balmer_2025}, this way being a main species present during planet formation \citep{Pacetti2022, elenia_2025}, or released via subsequent impact delivery mechanisms \citep{Madhusudhan2016}.
Given the role of the presence of atmospheric CO$_2$ on planetary chemistry, understanding whether interstellar CO$_2$ ice is inherited by the forming planet is key to unravel the habitability of Earth-like planets \citep{Watanabe_2024}. 
In turn, this depends on the so-called CO$_2$ snowline, which is where the CO$_2$ ice sublimates and becomes gaseous.
In this work, we aim to model the CO$_2$ adsorption on an atomistic model of an amorphous water ice surface and constrain the parameters that define the gas/ice partitioning in a protoplanetary disk, which defines the available CO$_2$ reservoirs during planet formation.

Species adsorption on ice surfaces is governed by its binding energy (BE), which impacts the adsorption, diffusion and desorption on the ice.
In this regard, BEs can be obtained experimentally by employing the Temperature Programmed Desorption (TPD) technique, a well-established technique to study adsorption phenomena on surfaces.
During TPD, the desorption of a given species from the surface is monitored by a mass spectrometer.
The BE is extracted from the Polanyi-Wigner equation \citep{Polanyi1925} using mathematical formalisms.
Several experimental studies report the BEs of astronomically relevant species: \citet{Collings2004,ULBRICHT2006,Edridge2013,He2016,Penteado2017,He2017, Kruczkiewicz2024}.
However, this experimental method presents several limitations: (i) the derived BEs depend on the adsorption regime (i.e., submonolayer, monolayer, or multilayer) as well as on the morphology and chemical composition of the substrate; (ii) the parameter from which BEs are inferred corresponds to the desorption enthalpy, which equates to the BE only in the absence of other activated processes; and (iii) its applicability is restricted to conditions involving continuous surface heating, thus failing to reproduce the cold environments of the interstellar medium (ISM) \citep{Ligterink_2025}.  
To alleviate such drawbacks, atomistic computational studies can be conducted, since they are free of these constrains. 
Previous theoretical investigations have been conducted to compute the BEs of O-,N-, and S-bearing species as well as COMs (complex organic molecules) by \citet{Ferrero2020,Martínez-Bachs_2024,Perrero2022, Kakkar_2025}, where the BEs were calculated on both crystalline and amorphous H$_2$O ices, providing a range of BEs for each species 
by identifying several binding sites on an amorphous water solid (AWS) structural model. 
This idea was extended into finding BE distributions, for several astrochemically relevant species \citep{Bovolenta2020,Germain2022,Tinacci2023, Bovolenta2022, Bariosco2024, Bariosco2025,Groyne2025,Bovolenta2025} using robust computational chemistry methodologies.

Efforts have also been made to compare computational and experimental determinations of binding energies by simulating TPD spectra from computational datasets \citep{Bariosco2024,Bariosco2025, Bovolenta2025}.
A notable source of uncertainty, however, arises from the approximation of the pre-exponential factor (prefactor) in the Polanyi–Wigner equation, which can substantially influence the BEs inferred from experimental TPD curves, as highlighted by \citet{Ferrero2022, Pantaleone2025}. 
Further limitations arise when comparing different coverage regimes. This issue has been addressed in \citet{Bovolenta2025}, who proposed a novel approach involving the simulation of different coverage regimes by adjusting the dataset fragments used to calculate the TPD spectrum.

In this work we aim to find the BE distribution of CO$_2$ on a realistic AWS atomistic model using the best accurate quantum chemical methodology with the large size of the considered ice model. 
The results are discussed based on the different energy contributions and the geometrical analysis, and then compared with experimental values using the novel coverage simulation approach and different prefactor models. 
The obtained distribution is also compared with values from previous theoretical studies.
Moreover, the effect of the presence of more than one CO$_2$ molecules on the surface on the BE is studied, testing the wetting and non-wetting properties of CO$_2$ on the AWS.
The influence of adsorption of CO$_2$ on AWS on the spectral features is also assessed. 
Finally, the impact of the BEs distribution and different prefactor models on the ice-gas partitioning in a protoplanetary disk is discussed.


\section{Methodology}
\label{method}

The procedure to obtain a BE distribution in this work is adopted from previous works by our group \citep{Germain2022,Tinacci2023,Bariosco2024,Bariosco2025}.

\subsection{BE definition}


BE is defined as the negative of the interaction energy ($\Delta$E), i.e. difference between the total energy of the adsorption complex (E$_{C}$) and the energy of each isolated components, namely, the grain (E$^{iso}_{grain}$) and the adsorbed molecule (E$^{iso}_{mol}$). 
This can be expressed in the following way:

\begin{equation}
    \mathrm{BE} = -\Delta E = E^{\text{iso}}_{\text{grain}} + E_{\text{mol}}^{\text{iso}} - E_C,
\end{equation}
\begin{equation}
    \Delta E = E_C - E^{\text{iso}}_{\text{grain}} - E_{\text{mol}}^{\text{iso}}.
\end{equation}

BE can be decomposed into two terms: (i) the electronic interaction energy (BE$_{e}$), which is corrected for the basis set superposition error (BSSE, calculated at the geometry of the complex), 
and (ii) the deformation energy ($\delta E_{\text{def}}$), which is the result of the geometrical change induced by the adsorption of the species on the grain surface. Finally, by subtracting the zero point energy ($\Delta$ZPE) inferred from harmonic frequency calculations, BE at 0K is obtained. That is:

\begin{equation}
    \mathrm{BE} = \mathrm{BE}_{e} - \delta E_{\text{def}} - \Delta \mathrm{ZPE}.
    \label{be_def}
\end{equation}
\begin{equation}
    \delta E_{\text{def}} = E_{\text{mol}}^{\text{C}} - E_{\text{mol}}^{\text{iso}} + E_{\text{grain}}^{\text{C}} - E_{\text{grain}}^{\text{iso}}
\end{equation}

\noindent
where the deformation energy ($\delta E_{\text{def}}$) is defined as the difference between the energy of the complex component ($E^{{C}}$) and fully relaxed at isolated geometry ($E^{{iso}}$), and $\Delta$ZPE is defined as

\begin{equation}
    \Delta\text{ZPE} = \text{ZPE}_{\text{C}} - \text{ZPE}^{\text{iso}}_{\text{mol}} - \text{ZPE}^{\text{iso}}_{\text{grain}}
    \label{ZPE}
\end{equation}

\subsection{Computational details}

\begin{figure}
    \centering
    \includegraphics[width=0.75\linewidth]{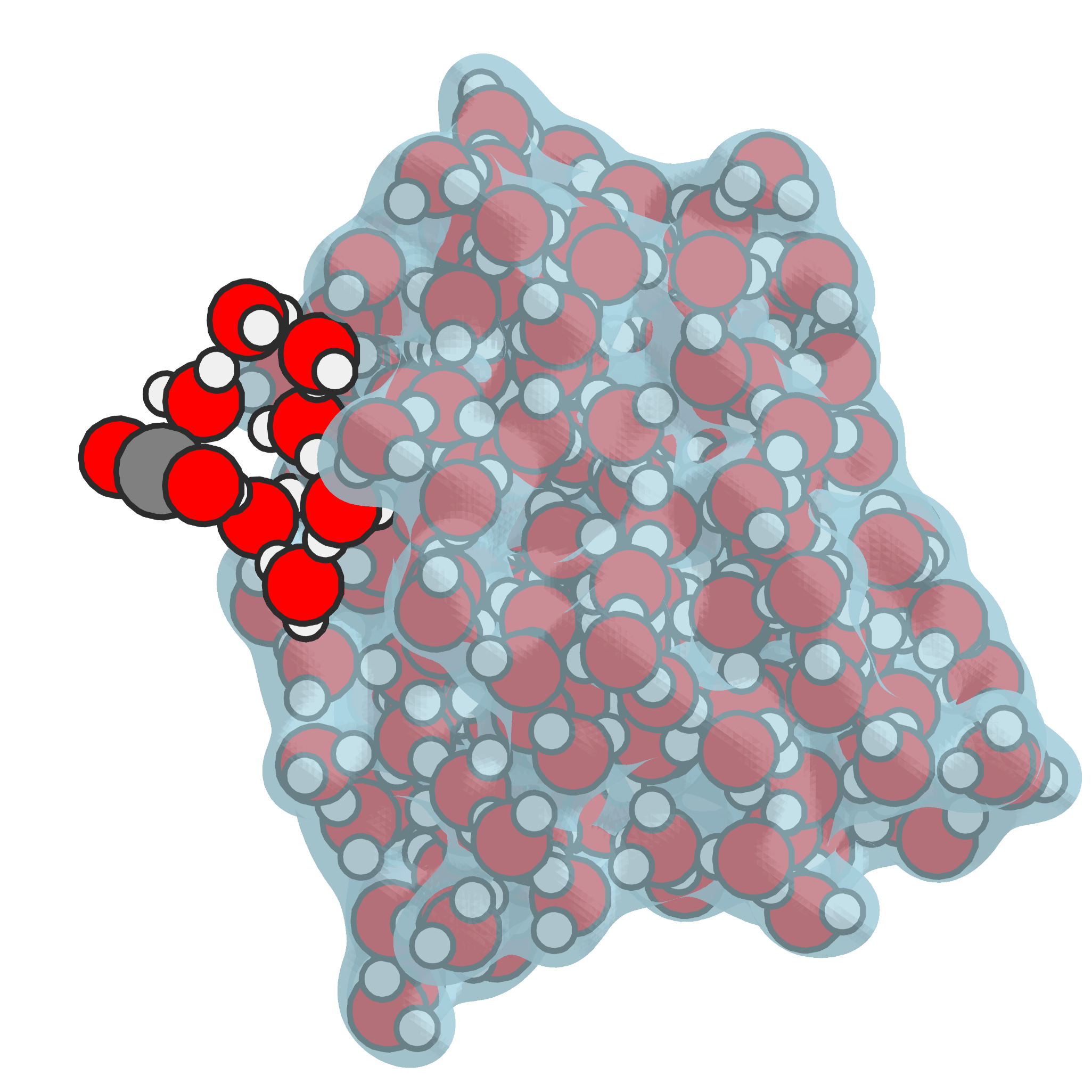}
    \caption{Atomistic cluster model for water ice used in this work, consisting of 200 water molecules. It also shows the ONIOM scheme used: the \textit{Real-Zone} is surrounded by the semi-transparent iso-surface, while the \textit{Model-Zone} is outside of the bubble created around the grain. Colour-coding: white, H atoms; grey, C atoms; red, O atoms.}
    \label{fig:oniom_scheme}
\end{figure}

\begin{figure*}[h!]
    \centering
    \includegraphics[width=0.90\linewidth]{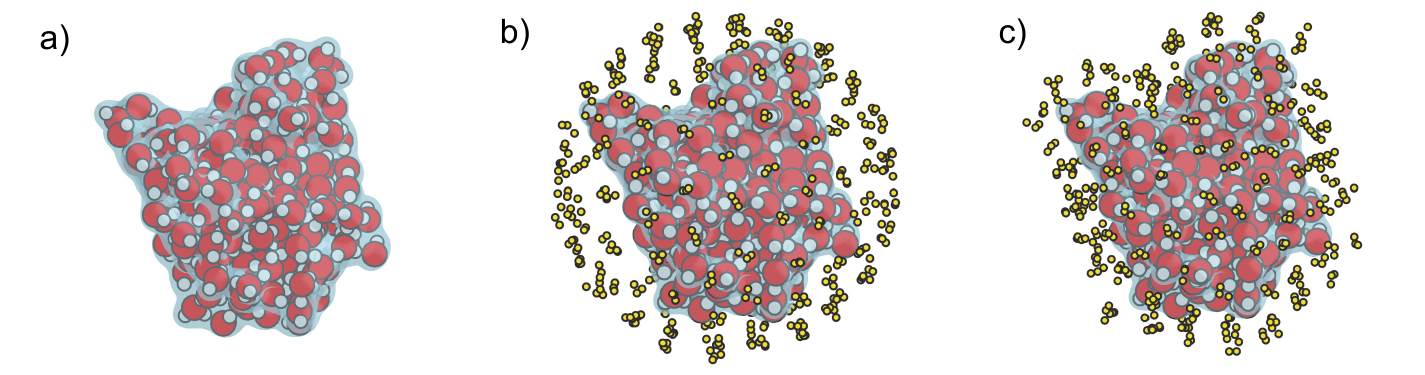}
    \caption{ Model of the water ice grain and its adsorption sites. a) 200 water molecules grain created with the ACO-FROST procedure \citep{Germain2022};
b) initial grid of 486 points reported as yellow spheres; c) final grid projected onto the grain at 2.5 \AA from the surface. Red and white spheres are oxygen and
hydrogen atoms, respectively, while yellow spheres are the centres of mass of the CO$_2$ molecule.}
    \label{fig:grain_sampl}
\end{figure*}

The structure of the water ice grain model along with the initial structures of the adsorbate-grain complex have been optimized using the semi-empirical GFN2-xTB methodology (the 2019 release) \citep{Grimme_2017,Bannwarth2019} to obtain preliminary BEs. 
Such initial structures were refined with the ONIOM method \citep{Mayhall_2010} (QM:QM2) using the ORCA (v.6.0.1) code \citep{RN269}. 
The ONIOM method is an established hybrid approach that enables the use of different full quantum mechanical, semi-empirical or classical mechanics methodologies on different parts of the system that gives reliable energies, geometries and harmonic frequencies at a lower computational cost. 
In the two-layer approach employed in this study the structure is divided into two substructures: 
i) the \textit{Real-Zone} consisting of the entire geometry and treated at the low-level of theory (QM2), and ii) the \textit{Model-Zone} consisting of the adsorbate and the water molecules within the 5 {\AA} of the barycentre of the adsorbate and treated at higher level of theory (QM). 
An example of the defined \textit{Model-Zone} and the grain is shown in Figure \ref{fig:oniom_scheme}.
The final ONIOM energy E(ONIOM) is obtained as:
\begin{equation}
    E\text{(ONIOM)} = E_{\text{Real}}(\text{QM2}) - E_{\text{Model}}(\text{QM2}) + E_{\text{Model}}(\text{QM})
\end{equation}
After a thorough benchmarking procedure (see Appendix) the B97-3c \citep{Brandenburg_2018} density functional theory (DFT) method was chosen for the optimization procedure, where the atoms in the Real zone are kept fixed, while the Model-Zone is embedded electrostatically as default setting in the ONIOM scheme. 
Additionally, an Extended Active Region setting was used to minimize the number of cases where imaginary frequencies appear. 
Both the self-consistent field (SCF) and the geometry optimizer criteria were set for a very tight convergence, i.e., total energy change < $2\times10^{-7}$Eh. 
For the integral evaluation, the grid was set to the highest density setting to ensure the highest possible accuracy of the numerical integration. 

After the optimization, the final energies were refined by performing single-point energy calculations at the DLPNO-CCSD(T) methodology \citep{Guo_2018} for the \textit{Model zone} in the ONIOM scheme. 
For these calculations, the selected primary basis set was the aug-cc-pVTZ \citep{Kendall_1992, Woon_1993}, while the auxiliary basis set used for the resolution of the identity approximation was aug-cc-pVTZ/C \citep{Weigend_2002}.
For the domain-based pair neutral orbitals, the truncation for the occupation number was set to $10^{-7}$, which allows to recover 99.99\% of the canonical coupled cluster correlation energy.

\subsection{Iced grain model and the procedure for computing the BEs}

The iced grain surface was simulated using a 200-water molecule cluster model, which was adopted from previous works \citep{Germain2022, Tinacci2023, Bariosco2024, Bariosco2025} to model an interstellar icy grain, see Figure \ref{fig:grain_sampl}a).
The sampling procedure used the ACO-FROST code \citep{Germain2022} analysing a grid of equally spaced 162 points projected onto the surface model. 
These points are representative of the initial adsorption positions of the species considered, in this case CO$_2$. 
In addition, the CO$_2$ molecule is rotated three times to facilitate the randomization of the sampling and eliminate potential bias for more energetically stable binding sites, thus resulting in 486 initial positions, see Figure \ref{fig:grain_sampl}b).
The code projects the species on the grain in a way that its barycentre is 2.5{\AA} away from the surface, see Figure \ref{fig:grain_sampl}c). 
This initial approach is followed by the following steps:

i) CO$_2$ molecule is optimized on the fixed surface of the grain at GFN2 level. 

ii) The optimized system from step (i) is further optimized at the GFN2 level, this time allowing the water molecules located within a 5{\AA} radius of the CO$_2$ barycentre to be optimized with the adsorbate.

iii) Following the optimization in step (ii), a geometrical check is performed to ensure that the number of water molecules within the 5{\AA} region around the CO$_2$ barycentre remains unchanged after relaxation. If any change is detected, steps (ii) and (iii) are repeated iteratively until convergence is achieved with respect to the number of molecules included in the optimization region.

iv) After convergence is achieved in (iii), the (QM:QM2) ONIOM scheme is applied on the resulting structures. 
Similarly to the GFN2 procedure, the waters present within the 5{\AA} radius from the adsorbate are included in the \textit{Model-Zone} and are free to relax. 
At the end of the optimization procedure at the B97-3c level, convergence check on the \textit{Model-Zone} was done as in (iii) allowing for one water molecule difference.

v) Frequencies at the harmonic approximation are computed for the \textit{Model-Zone}, while the remaining atoms are kept fixed.

vi) After the optimization, the molecule is removed from the surface (resembling its desorption) and the grain model is again reoptimized, so that the $E_{\text{grain}}^{\text{iso}}$ could be defined according to the TPD approach.

vii) Finally, the energies were refined through single point calculations using the DLPNO-CCSD(T):GFN2 ONIOM scheme, for both the bare grain and the complex structures.

\subsection{Procedure to obtain unique BE sites}{\label{procedure_be}}

A pruning procedure was implemented in order to avoid redundancies in the BE sites and ensure a uniform sampling of the BE in the automatic procedure. 
The benchmark (see Appendix) revealed that low-cost methods including GFN2 find two minima for the H$_2$O$\cdots$CO$_2$ complex, while the more accurate methods (such as DFT and CCSD(T)) identify only one. Therefore, a double pruning procedure was employed, where the geometries were pruned both at GFN2 level and after ONIOM optimization at B97-3c level.
As shown in \citet{Bariosco2024} and discussed in \citet{Tinacci2023}, this procedure significantly reduces the computational effort. 
Redundant structures were evaluated in the following way: 
i) all combinations where the model zone contained the same amount of atoms were considered;
ii) for these combinations, the geometries were aligned and their RMSD was computed along with the absolute value of the energy difference $|E_{C_{ij}}| = |E_{C_{i}} - E_{C_{j}}|$
iii) pruning criteria were set based on the correlation between RMSD and $|E_{C_{ij}}|$, and were different for the pruning at DFT and GFN2 levels. Details can be found in the Appendix.

\subsection{Desorption rate prefactor}
\label{sec: prefactor}

Previous studies \citep{Ferrero2022,Minissale2022, Ceccarelli2023Organic,Minissale2023, Tinacci2023, Pantaleone2025, Bariosco2025} suggest that some focus should be given to the modelling of the pre-exponential factor (prefactor, $\nu$(T)) value, since it directly affects the desorption rate ($k_{des}$), that is:

\begin{equation}
\label{eq:desorption}
    k_{des} = \nu(T)\exp{ \left( \frac{-E_{des}}{T} \right) }
\end{equation}
Based on \citet{Pantaleone2025}, where the applicability of different models on soft surfaces (such as amorphous water ice) was discussed, we adopt two methods to compute the prefactor: (i) $v_{Tait}$, proposed by \citet{Tait2005, Tait2006}, and (ii) $v_{Campbell}$, proposed by \citet{Campbell2012, Minissale2022}, hereafter referred to as Tait and Campbell prefactors, respectively. 
Details on the TPD spectra simulation and the prefactors can be found in the Appendix.

\subsection{Multi-molecule adsorption regime}


For the adsorption of multiple CO$_2$  molecules, two scenarios were established representing the two extreme cases: one in which the CO$_2$ molecules are as far apart as possible, the other in which the CO$_2$ are in close proximity forming islands. This was done by modifying the ACO-FROST procedure.
In the first case, the aim was to maximize the distance between molecules. 
In order to do that, the procedure identified the grid point furthest from any existing CO$_2$ on the grain surface. 
Once this optimal spot was found, the new molecule was randomly rotated and placed on the grain.
Then the geometry and final BE are obtained according to the procedure discussed in Section \ref{procedure_be}.
In the second case, the goal was to create a dense CO$_2$ cluster on the water grain. This was done by considering an adsorption spot randomly from the five points closest to previous existing CO$_2$ molecules, where the new CO$_2$ molecule is randomly rotated before being adsorbed at 2.5 \AA distance in the selected site. 
To allow for vertical stacking, the grid accounts for previously adsorbed molecules.
The procedure to obtain the BE was the following:

i) steps i) and iii) from Section \ref{procedure_be} were followed.

ii) the (QM:QM2) ONIOM scheme was applied, which is always defined as the molecules within 5 \AA radius from the most recently adsorbed molecule. The \textit{Model-Zone} was optimized using B97-3c methodology and the convergence check described in Section \ref{procedure_be} in point iv) was done.

iii) Frequencies at the harmonic approximation were computed for the \textit{Model-Zone}, while the remaining atoms were kept fixed.

vi) after the optimization, the most recently adsorbed molecule was desorbed from the surface and the grain model was again reoptimized, such that the $E_{\text{grain}}^{\text{iso}}$ could be defined according to the TPD approach.

vii) finally, the energies were refined through single point calculations using the DLPNO-CCSD(T):GFN2 ONIOM scheme, for both the bare grain and the complex structures.

For both cases, a total of 10 CO$_2$ molecules were adsorbed one-by-one on the surface starting from the same initial structure with a single CO$_2$ molecule.

\section{Results}

In this section, the obtained BE distribution will be presented along with the energetical decomposition and geometrical analysis. 
The results of the multi-molecule adsorption are also introduced.


\subsection{CO$_2$ BE distribution}
\label{res:single_be}

\begin{table*}[h!] 
\caption{Mean BE value, prefactor evaluated at the peak temperature ($\nu$(T$_\mathrm{{peak}}$)), associated T$_\mathrm{{peak}}$ and relative population for each bin of the histogram}
\label{tab:binned_prefactor}
\centering
\begin{tabular*}{\linewidth}{@{\extracolsep{\fill}} lccccr @{}}
\hline\hline 
\text{Mean } $\text{BE}$ & \text{Fraction of the ice} & \multicolumn{2}{c}{{Tait Model}} & \multicolumn{2}{c}{{Campbell Model}} \\
\hline 
\text{(K)} & \text{(\%)} & $\nu(T_{\text{peak}})$ $(\text{s}^{-1})$ & $T_{\text{peak}}$ (K) & $\nu(T_{\text{peak}})$ $(\text{s}^{-1})$ & $T_{\text{peak}}$ (K) \\
\hline 
$1205.1$ & $4.92$ & $1.898 \times 10^{15}$ & $32.5$ & $6.128 \times 10^{13}$ & $36.0$ \\
$1519.8$ & $15.57$ & $4.100 \times 10^{15}$ & $40.5$ & $7.976 \times 10^{13}$ & $44.5$ \\
$1743.6$ & $21.31$ & $6.162 \times 10^{15}$ & $45.5$ & $9.450 \times 10^{13}$ & $51.0$ \\
$1997.1$ & $13.12$ & $9.507 \times 10^{15}$ & $51.5$ & $1.097 \times 10^{14}$ & $57.5$ \\
$2206.9$ & $18.03$ & $1.315 \times 10^{16}$ & $56.5$ & $1.241 \times 10^{14}$ & $63.5$ \\
$2449.1$ & $17.21$ & $1.820 \times 10^{16}$ & $62.0$ & $1.401 \times 10^{14}$ & $70.0$ \\
$2667.9$ & $6.56$ & $2.388 \times 10^{16}$ & $67.0$ & $1.539 \times 10^{14}$ & $75.5$ \\
$2916.6$ & $3.28$ & $3.147 \times 10^{16}$ & $72.5$ & $1.719 \times 10^{14}$ & $82.5$ \\
\hline 
\end{tabular*}
\end{table*}

The final BE distribution of CO$_2$ on the water ice model was obtained from 122 unique sites that remained after the last pruning procedure. 
All of them were PES minima, as revealed by frequency analyses (no imaginary frequencies were found). 
The final number of the unique sites identified in this study is lower than those reported in literature, which is due to the fact that the CO$_2$ molecule is highly symmetrical, thus the probability that the procedure will produce redundant geometries becomes increased.
After the energy refinement at the DLPNO-CCSD(T) level, data was organized in the form of a histogram using the Knuth binning method \citet{KNUTH2019}. This method allows for the determination of the appropriate number of bins using the Bayesian approach, where the posterior probability of the binning model is maximized given the observed data. 

\begin{figure}[h!]
    \centering
    \includegraphics[width=1\linewidth]{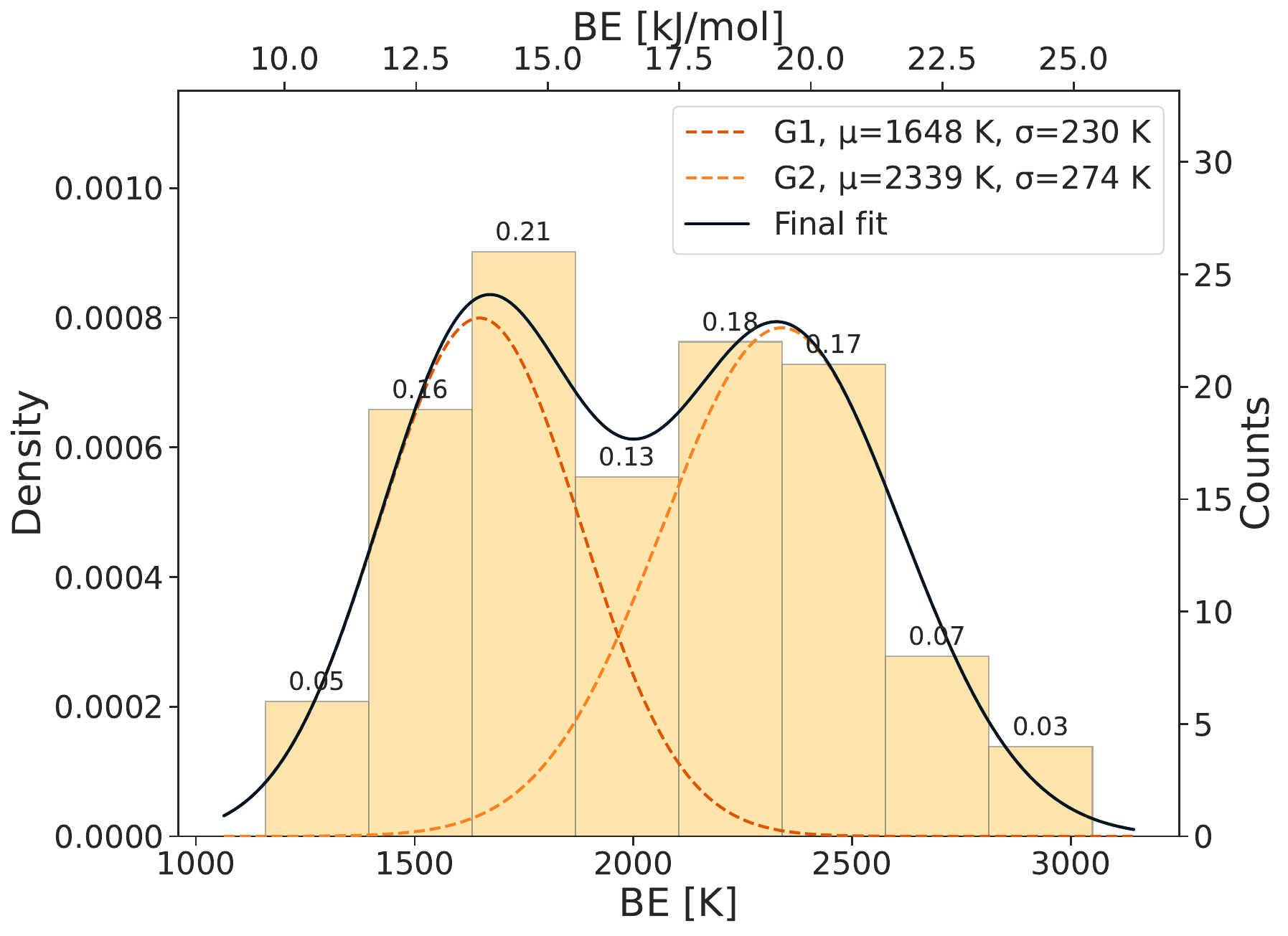}
    \caption{Plot showing binned BE values at DLPNO-
CCSD(T)/aug-cc-pVTZ//B97-3c level, the dotted lines represent the contribution of each Gaussian to the overall bimodal Gaussian distribution which is shown using a black line.}
    \label{fig:BE_distr}
\end{figure}

The distribution parameters were obtained using the bootstrapping procedure.
Moreover, a likelihood-ratio test (LRT) was performed to verify that the data indeed follows a double Gaussian function vs. a single Gaussian function.
The single Gaussian function was defined as:

\begin{equation}
G(x) = \frac{1}{\sigma \sqrt{2 \pi}} \exp \left( -\frac{(x - \mu)^2}{2 \sigma^2} \right)
\end{equation}

\noindent where $\mu$ is the expected value and $\sigma^2$ is the variance. 
The bimodal Gaussian function was defined as:

\begin{equation}
    F(x) = w \cdot G_1 + (1 - w ) \cdot G_2
    \label{Gausian}
\end{equation}

\noindent where G$_1$ and G$_2$ are separate Gaussian functions defined in Equation \ref{Gausian} and $w$ is a scaling parameter.

The defined Gaussian functions are defined as probability density functions (PDFs), and therefore the data has been converted from counts to probability density.  
The bootstrapped parameters were for the first Gaussian function G$_1$: $\mu_1$, $\sigma_1$, and for the second one G$_2$: $\mu_2$, $\sigma_2$, and the scaling parameter for the sum of the two Gaussians was $w$. 
The details of the bootstrapping procedure are given in the Appendix.
The resulting parameters obtained adopting the two Gaussian models are: 
\[
\begin{aligned}
\text{G}_1: \quad & \mu_1 = 1648^{+190}_{-109}, & \sigma_1 = 229^{+99}_{-81}, \\
\text{G}_2: \quad & \mu_2 = 2339^{+185}_{-181}, & \sigma_2 = 274^{+0.288}_{-0.226}, \\
& w = 0.46^{+0.274}_{-0.221} 
\end{aligned}
\]

The LRT showed a preference towards the double Gaussian function with $p<0.05$.
The histogram of the data and the bootstrapped distribution is shown in Figure \ref{fig:BE_distr}.
In Table \ref{tab:binned_prefactor}, the population of each bin along with its mean BE value is reported.
Both types of prefactor for each bin at the desorption peak (T$_{peak}$) were calculated numerically based on the Polanyi-Wigner equation \citep{Polanyi1925}, as in previous works done by \citet{Bariosco2024,Bariosco2025}. 
The desorption peak temperature (T$_{peak}$) is defined as the temperature at which the desorption rate reaches a maximum. 

\subsection{BE decomposition and geometrical analysis}

\begin{figure*}[ht!]
    \centering
    \includegraphics[width=1.0\linewidth]{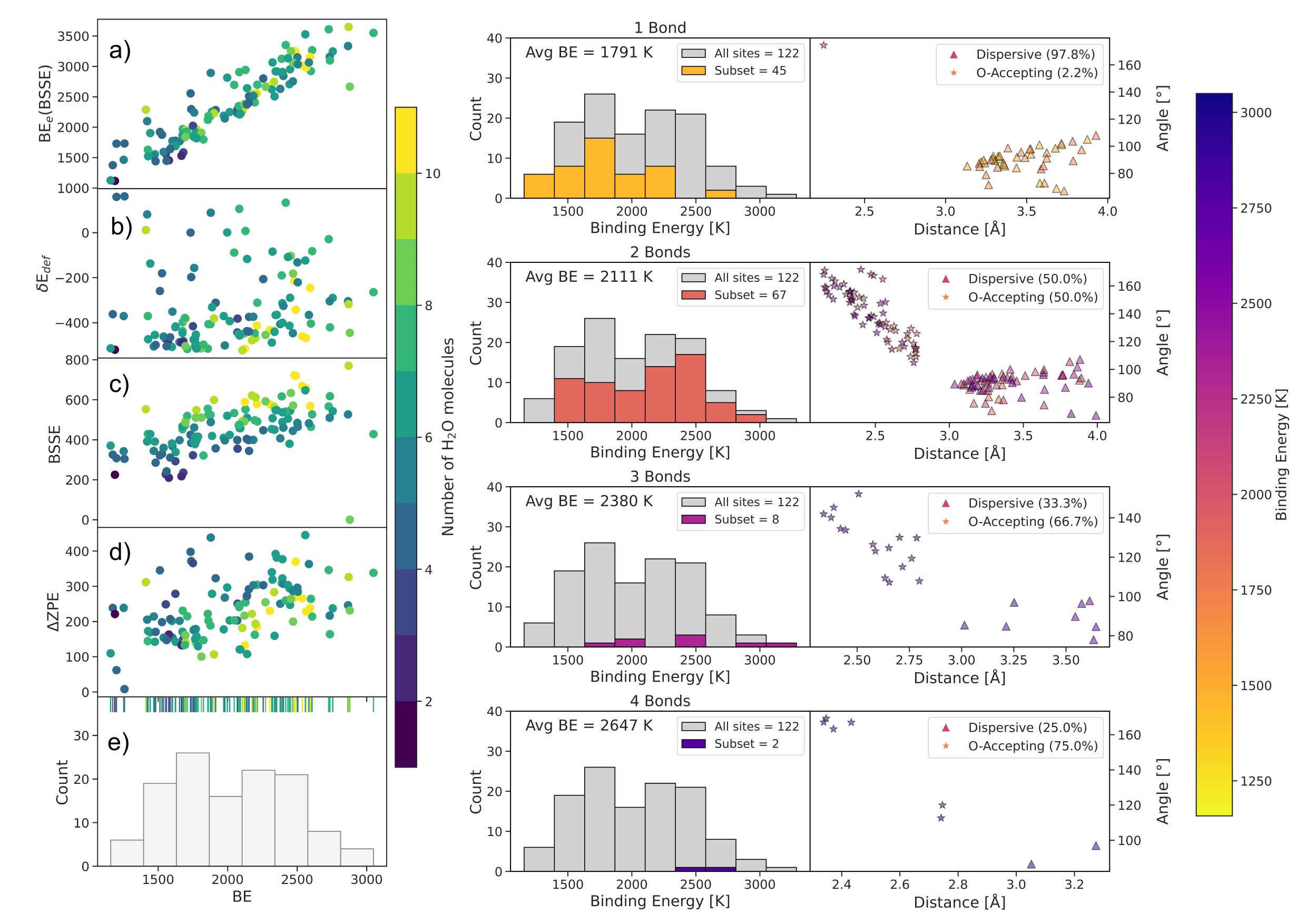}
    \caption{Left panel: plots showing the correlation between the BE and several contributions to the BE, plot a): electronic BE which is not corrected for the BSSE (BE$_e$(BSSE), plot b) deformation energy ($\delta$E$_{def}$), plot c) BSSE correction, plot d) $\Delta$ZPE correction (see equation \ref{be_def}), and plot e) shows the BE distribution. BEs showed on the X-axis are in Kelvin computed at DLPNO-CCSD(T)/aug-cc- pVTZ//B97-3c level. The colour map on the right side shows the number of H$_2$O molecules included in the Model Zone. Right panel: Plot showing the geometry decomposition, the plots on the left show the BEs corresponding to the selected geometry types along with mean BE for the subset and the subset size, and the plots on the right show the geometry parameters: interaction distance and the angle. From the top the criteria are: 1 bond formed with the surface, 2 bonds, 3 bonds and 4 bonds.}
    \label{fig:correlation_plot}
\end{figure*}

The computed values of BE can be decomposed into different energy contributions: the deformation energy, the BSSE and $\Delta$ZPE. 
In previous theoretical works (e.g., \citet{Martínez-Bachs_2024, Ferrero2020, Perrero2022, Bulik2025}), the dispersion contribution was quantified. 
However, in this case, the dispersion is only accounted for as a separate term during the B97-3c optimization, while for the DLPNO-CCSD(T) dispersion it is not calculated as a separate term.
The different contributions are colour coded with the number of H$_2$O molecules included in the \textit{Model-zone}. They are shown in the left panel of Figure \ref{fig:correlation_plot}.
The a) plot represents the electronic energy, described with BE$_e$(BSSE), which is the binding energy of the CO$_2$ molecule and the grain that has not been optimized after the molecule desorption.
The following contributions show the necessary corrections applied to the BE$_e$(BSSE) to obtain the final BE.
The plot b), marked with $\delta$E$_{def}$, represents the deformation energy occurring when the surface is reoptimized after the desorption of the molecule. 
Contrary to previous works \citep{Germain2022,Tinacci2023,Bariosco2024,Bariosco2025}, we found negative contributions to the deformation energy, which make the BE higher since the adsorption of the species on the grain stabilizes the surface. 
There is also no clear correlation between the $\delta$E$_{def}$ and the BE nor the size of the \textit{Model-zone}. 
The c) plot shows the contribution of the BSSE to the total BE. 
The correlation exhibits a trend where the higher the BE, the higher the error, and the larger the \textit{Model-zone}, the higher the error. 
This is an expected behaviour, which arises from the definition of the BSSE.
The overlapping functions of the BSSE will contribute more energy if more atoms are involved in the \textit{Model-zone}.
The d) plot shows the $\Delta$ZPE contribution to the BE, which is not clearly correlated with the \textit{Model-zone} size nor the BE value.

A geometry analysis was performed, where two types of interactions were defined: (i) H-bonding, where one of the two CO$_2$ oxygen atoms acts as the H-bond acceptor, and (ii) dispersion interactions. 
The criteria for the H-bond were: the bond distance HO-H $\cdots$ O-CO should lie between 1.7\AA and 2.8\AA, the bond angle defined as the angle marked by donor oxygen, donor hydrogen from the water molecule and acceptor oxygen from the CO$_2$, should be expected from 100$\mathring{}$ to 180$\mathring{}$. The dispersion interaction was defined as a distance from the surface along with with the angle between CO$_2$ and the surface, the selected range for the distances being 2.8-4 \AA, and for the angles 65-115$\mathring{}$.

The right panel of Figure \ref{fig:correlation_plot} shows the results of the geometrical analysis. 
We defined four situations based on the total number of bonds formed with the surface according to the above definitions.
The plots on the left show the histogram of all BEs values along with the highlighted BEs that were put each category.
The plots on the right show the distance and angle of each bond for each geometry, which is colour coded with the corresponding BE.
Geometries where only one bond has been identified are found on the lower end of the BE distribution, with an average BE of 1791 K. 
This subset consists of 45 structures, including one instance CO$_2$ acting as an H-bond acceptor, for the remaining 44 cases the sole interaction is through dispersion. 
In the two bond situation, the largest subset consists of 67 geometries, where the average BE is 2111 K. 
In this case, the CO$_2$ molecule is interacting with the surface with a dispersion interaction and through one H-bond. 
This is the most common configuration, since it covers almost the entire range of the distribution of the BEs (excluding the lowest end).
The colour map in this plot shows that for the O-accepting bond, the higher (lower) the angle (distance), the higher the BE, while for the dispersion there is clear clustering of distances around 3.00 - 3.25 \AA.
In the three bonds situation (namely, two H-bonds and 1 dispersion interaction) has been identified for 8 geometries, with an average BE of 2380K.
Two cases have been identified as four bond situation, where one of the two oxygen atoms is able to form a 2 H-bonds: one weaker and one stronger. The BEs for these two cases is centred around 2647 K, which is close to the mean values reported in \citet{Ferrero2020} (2629 K) and \citet{Bulik2025} (2888 K).
These are not the highest BEs in the distribution, which suggests that the second H-bond is not essential for geometry stabilization in the case of CO$_2$.

\subsection{Multi-molecule adsorption and BEs}


The intermolecular interactions of CO$_2$ on the iced grain were evaluated by adsorbing multiple CO$_2$ on the surface via two different scenarios.
Firstly, the molecules were adsorbed in a way that maximized the distance between them on the grain, hereafter referred to as the \textit{far scenario}. This probed the long-distance interactions between the species, aimed at bridging the gap between cluster and periodic surface modelling.
This approach simulated the non-wetting behaviour of the species on the surface.
In the second approach, a clustering of CO$_2$ molecules was emulated by adsorbing the molecules close to each other, which imitated the wetting behaviour of the species (hereafter referred to as the \textit{cluster scenario}).
The resulting BEs are shown in Table \ref{tab:multipleBEs}.
For both approaches, the starting geometry corresponds to the geometry that represents the mean of the second peak of the bimodal Gaussian distribution from Section \ref{res:single_be}.
For the \textit{far regime}, the BEs cover the entire range of BEs set by the single-molecule adsorption in Section \ref{res:single_be}.
This indicates that there is no intermolecular contribution (negative nor positive) from additional CO$_2$ present on the iced grain, when adsorbed at a given distance.
However, when the clustering of the CO$_2$ is simulated, a clear trend appears: when there are fewer CO$_2$ molecules (up to 6) the BEs fall into the higher peak of the BE distribution, while when the number of the CO$_2$ molecules increases (above 6) the BEs fall closer to the lower peak of the BE distribution. This behaviour reflects changes in the local interaction environment of CO2. As the CO$_2$ amounts increase, CO$_2$--CO$_2$ interactions become more prevalent than CO$_2$-–H$_2$O interactions. Because CO$_2$--CO$_2$ interactions are intrinsically weaker than CO$_2$-–H$_2$O interactions, the overall binding energies decrease accordingly upon cluster adsorption of more than 6 CO$_2$ molecules. The increasing number of CO$_2$ molecules causes the BEs to plateau around 1550-1630K, which is caused by dispersion interaction. 
The CO$_2$ molecules relax into the geometry where they lay parallel to each other, which is favoured when dispersion is the dominant interaction.

\begin{figure}
    \centering
    \includegraphics[width=0.95\linewidth]{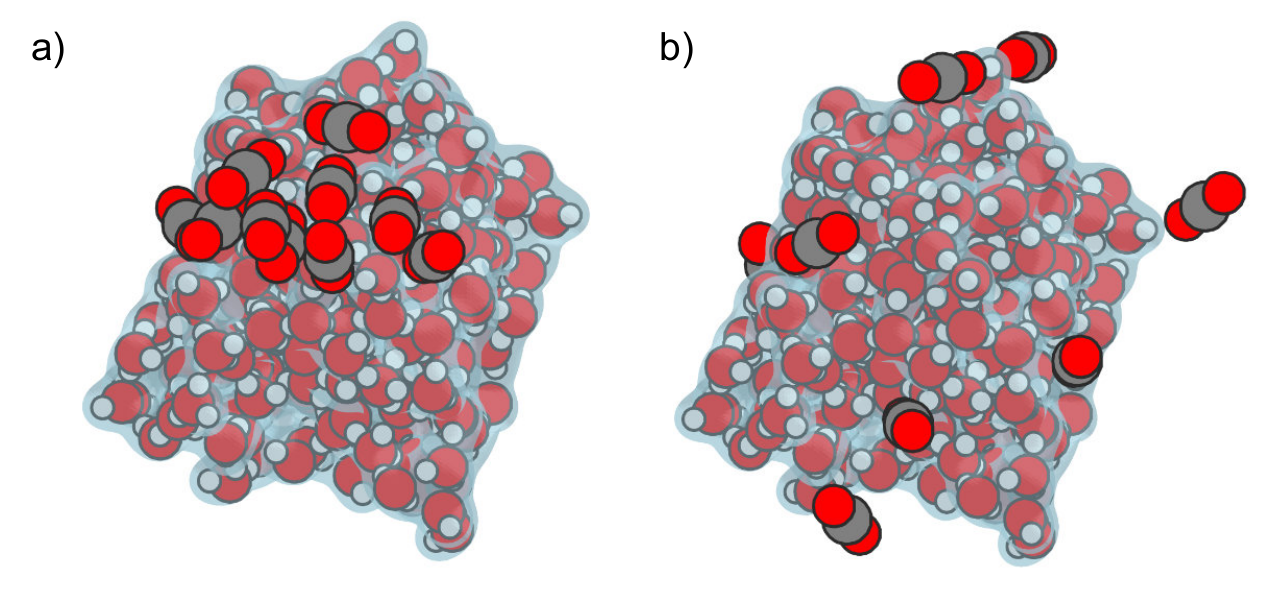}
    \caption{Figure a) shows the final geometry of the \textit{cluster scenario}, after all 10 CO$_2$ molecules have been adsorbed, Figure b) shows the final geometry of the \textit{far scenario}, after all 10 CO$_2$ molecules have been adsorbed.  Colour-coding: white, H atoms; grey, C atoms; red, O atoms.}
    \label{fig:placeholder}
\end{figure}


\begin{table}[]
    \centering
    \begin{tabular*}{\linewidth}{@{\extracolsep{\fill}} lcr @{}} \hline \hline
         n$^{o}$ CO$_2$ & far BE (K) & cluster BE (K) \\ \hline
         2 & 2504.2 & 2448.6\\
         3 & 1616.6 & 2311.7\\
         4 & 2716.9 & 2281.2\\
         5 & 1675.4 & 2471.6\\
         6 & 2386.0 & 2958.5\\
         7 & 2510.3 & 1560.5\\
         8 & 1242.2 & 1599.2\\
         9 & 2184.1 & 1628.7\\
         10 & 1672.8 & 1519.2 \\ \hline 
    \end{tabular*}
    \caption{The BEs in the multi-molecule approach are reported at DLPNO-CCSD(T)/aug-cc-pVTZ//B97-3c level of theory. The first column reports how many CO$_2$ molecules are present on the iced grain; in the second and third column, the BEs computed for the \textit{far scenario} and the \textit{cluster scenario} are reported, respectively.}
    \label{tab:multipleBEs}
\end{table}

\section{Discussion}

\subsection{Comparison with experimental BEs}\label{experiment}

\begin{table*}[ht!]
\caption{\label{table:tpd_experimets}Values of the CO$_2$ BE in K  and prefactors as measured in literature experiments using different substrates, ice samples and methods.}
\centering
\begin{tabular*}{\linewidth}{@{\extracolsep{\fill}} lccccc @{}}
\hline\hline
BE (K) & TPD peak (K) & Prefactor & Substrate & Coverage & Ref.\\ \hline
2886.5 $\pm$ 240.5 & 82 & $6\times10^{14\pm1}$ \tablefootmark{a} & HOPG & Monolayer & \tablefootmark{d}\\
2267 $\pm$ 71 & 72 & $9.3\times10^{26}$ \tablefootmark{b} & ASW & Submonolayer & \tablefootmark{e}\\
2105 $\pm$ 902 & 90 & $2.7\times10^{7\pm2}$ \tablefootmark{a} & ASW & Monolayer & \tablefootmark{f}\\
1100 & 87.3 & $5.2^{+4.3}_{-1.6}$ \tablefootmark{b} & Ice mixture on KBr \tablefootmark{c} & Multilayer & \tablefootmark{g}\\
2250 $\pm$ 20 & 78, 83 & $10^{12}$  \tablefootmark{a} & np-ASW & Submonolayer & \tablefootmark{h}\\
2325 $\pm$ 95 & 78 & $\sim$ & ASW & $\sim$ & \tablefootmark{i}\\
2814 $\pm$ 24 & 80 & $10^{13}$ \tablefootmark{a} & HAC(2) & Monolayer & \tablefootmark{j}\\
2750 $\pm$ 253 & 77 & $2.2\pm2.4\times10^{14}$ \tablefootmark{k} & Gold & Multilayer & \tablefootmark{k}\\
\hline
\end{tabular*}
\tablefoot{
\tablefoottext{a}{First-order desorption (s$^{-1}$)}
\tablefoottext{b}{Zero-order desorption (mol cm$^{-2}$s$^{-1}$)}
\tablefoottext{c}{Pre-cometary ice mixture: H$_2$O, CO, CO$_2$, NH$_3$, CH$_3$OH}
\tablefoottext{d}{\citet{ULBRICHT2006}}
\tablefoottext{e}{\citet{Noble2012}}
\tablefoottext{f}{ \citet{Edridge2013} }
\tablefoottext{g}{ \citet{Martin2014}}
\tablefoottext{h}{ \citet{He2017}}
\tablefoottext{i}{\citet{Penteado2017, Collings2004}}
\tablefoottext{j}{\citet{mate2019} }
\tablefoottext{k}{zero-order desorption (ML s$^{-1}$), \citet{Kruczkiewicz2024}}
}
\end{table*}

\begin{table}[h!] 
\caption{BE groups and their relative percentage in the distribution.}
\label{table:BE_groups}
\centering
\begin{tabular*}{\linewidth}{@{\extracolsep{\fill}} lcc @{}}
\hline\hline
Group & BE range (K) & Percentage (\%) \\ 
\hline
Very Low (VL) & 1158 -- 1412 & 5.74 \\ 
Low (L) & 1412 -- 1624 & 13.11 \\ 
Medium (M) & 1624 -- 2446 & 62.30 \\ 
High (H) & 2446 -- 2735 & 13.11 \\ 
Very High (VH) & 2735 -- 3048 & 5.74 \\ 
\hline
\end{tabular*}
\end{table}

\begin{figure}
    \centering
    \includegraphics[width=0.95\linewidth]{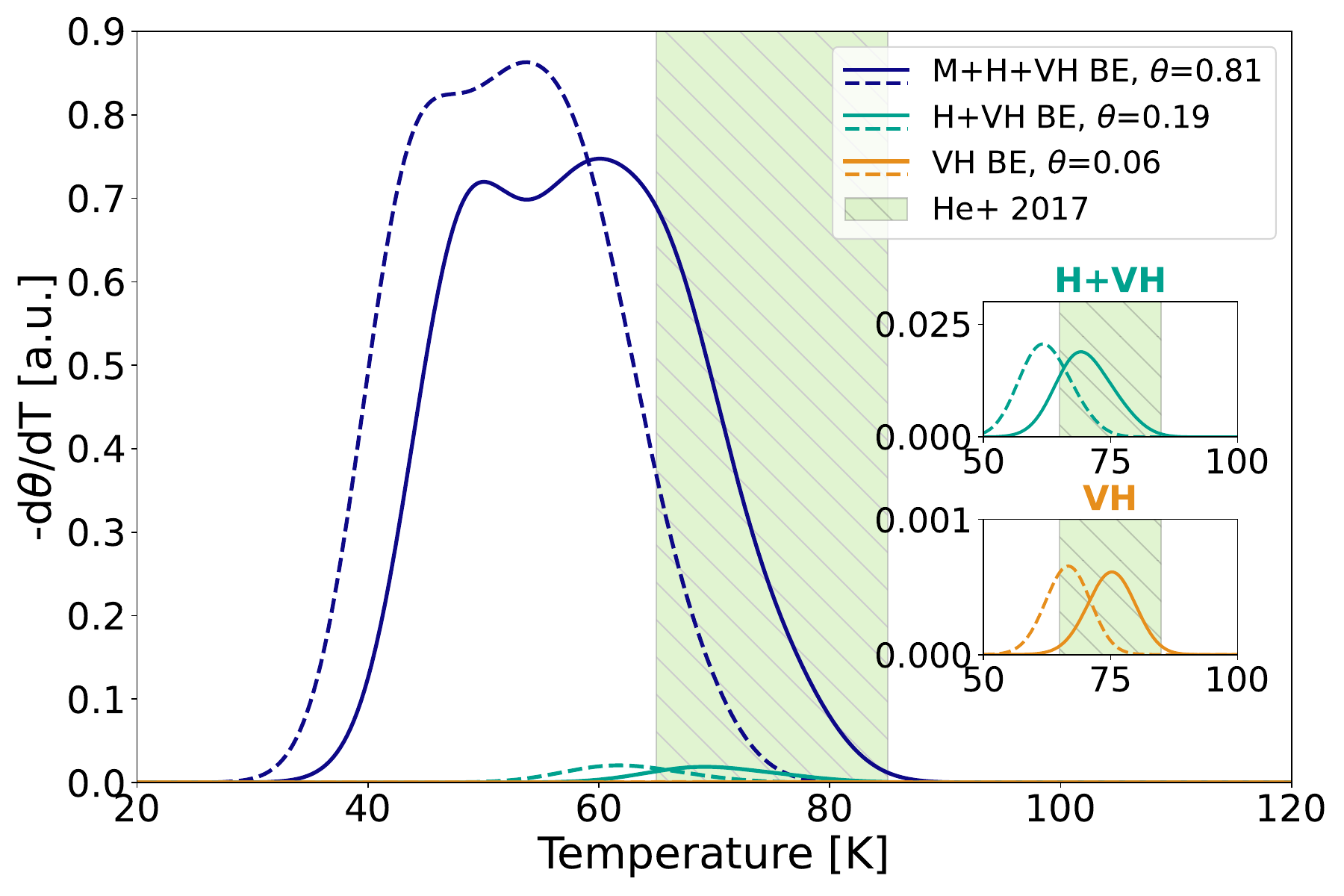}\\[1em]
    \includegraphics[width=0.95\linewidth]{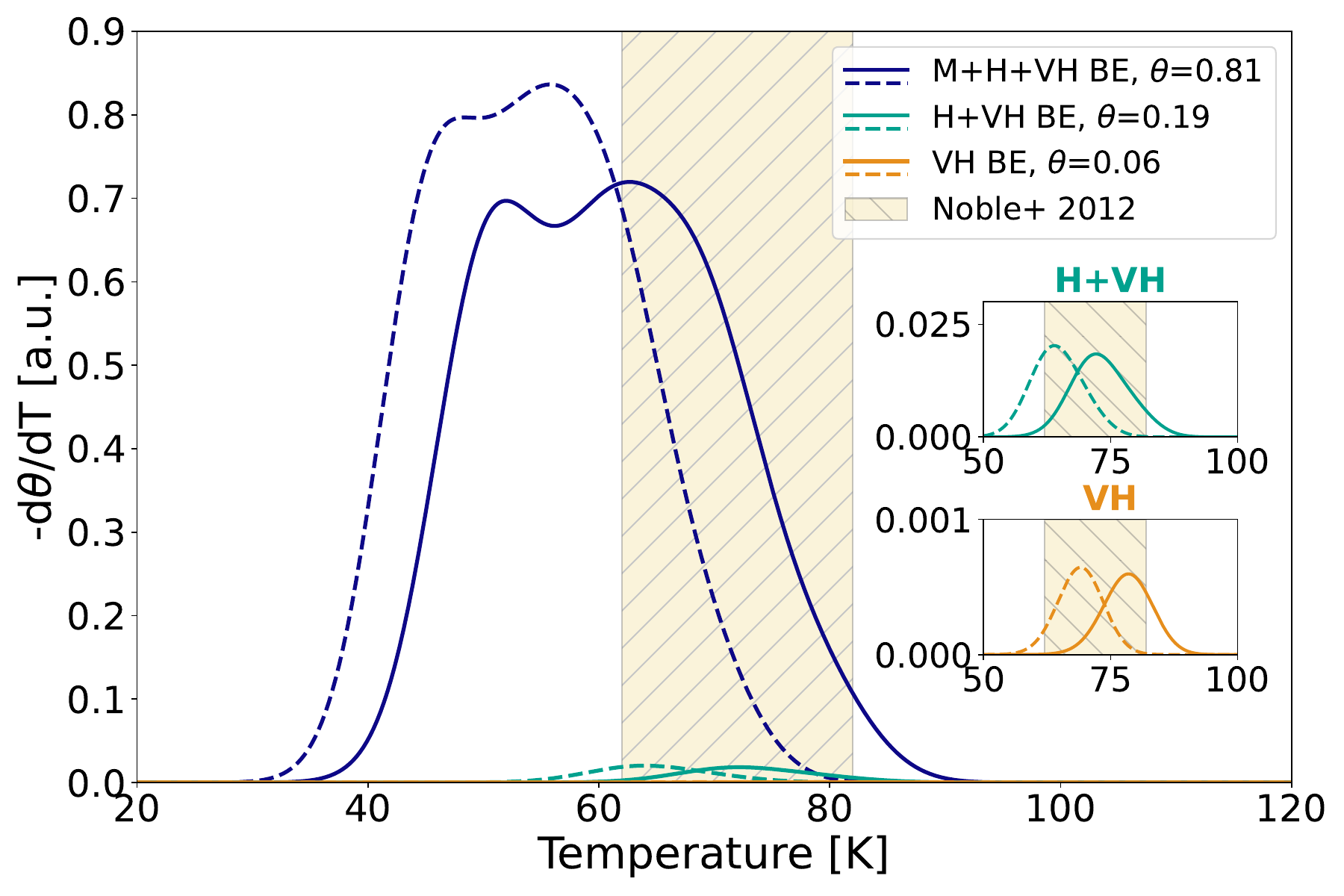}
    \caption{TPD curve simulation using Tait (dashed line) and Campbell (solid line) prefactors and simulation of different coverage regimes, top plot shows the comparison with the results of \citet{He2017} with $\beta$=2 K min$^{-1}$, $\theta$ = 0.6ML, the bottom plot shows the comparison with the results of \citet{Noble2012} with $\beta$=10 K min$^{-1}$. The insets in the plots show the desorption peaks of lower coverages that are out of scale compared with the largest coverage regime. The abbreviations: "M+H+VH" refers to medium, high and very high, "H+VH" refers to high and very high, "VH" refers to very high BE distribution parts.}
    \label{fig:tpd_spectra}
\end{figure}

In this section, our results are compared with those from previous experimental studies. 
The results of selected studies are shown in Table \ref{table:tpd_experimets}.
There are several factors that can impact the desorption energy derived in such experiments; for instance, the heating rate ($\beta$), which can affect the position of the peak in the spectrum.
The type of the substrate can also influence the BE value, as the binding affinity changes depending on the chemical properties of the material.
Moreover, the layering regime exerts a significant effect on the desorption behaviour.
The different regimes are: i) monolayer, which is defined as a single, closely packed layer of atoms, molecules, or cells that is only one unit thick; ii) sub-monolayer, where not all sites of the monolayer are occupied by the adsorbate; iii) multilayer, where more than one monolayer is adsorbed.
In the multilayer regime, the adsorbed species primarily interact with each other rather than with the substrate, and the derived desorption parameters therefore reflect intermolecular interactions between the same species. In contrast, in the monolayer and sub-monolayer regimes, the interaction between the adsorbate and the substrate becomes dominant, strongly affecting the desorption parameters \citep{Dombrowski_2021}.
To tackle this issue, the approach of \citet{Bovolenta2025} was adopted here. 
The BE distribution dataset was divided based on the adsorption strength into 5 bins, that were set based on the standard deviation and the 68-95-99 rule, the percentage of the distribution contained in each bin and the BE range are shown in Table \ref{table:BE_groups}.
The TPD spectra are simulated by emulating the sub-monolayer regime by grouping the subsets in the following way: (i) highest coverage BE range includes medium (M) + high (H) + very high (VH) subsets (see Table \ref{table:BE_groups}), (ii) medium coverage includes H + VH subsets, (iii) lowest coverage includes just the VH subset.
Following the approach in \citet{Bovolenta2025}, the coverage ($\theta$) is defined by the percentage of the values from the distribution taken into account when simulating the TPD curve.
Two prefactor models were employed, namely Tait and Campbell, (see Section \ref{sec: prefactor} and the Appendix).
The final result of the simulation is shown in Figure \ref{fig:tpd_spectra}, where it is superimposed with the range of the desorption peak obtained by \citet{He2017} and \citet{Noble2012}.
The heating ramp $\beta$ was selected according to the experimental set up of each of the selected studies, which are reported in the fourth column of Table \ref{table:tpd_experimets}.
Both experimental studies were performed in the sub-monolayer regime. 
For both comparisons, the lowest simulated coverages ("VH BE" regime) correspond better with the experimental data.

In the case of \citet{He2017}, the peak simulated with the Campbell prefactor falls within the range of the experimental peak, which is around 78K.
For \citet{Noble2012}, the experimental range is best represented by a peak that falls in between the two modelled peaks, close to 72K.
Based on the comparison, it can be notes that the Campbell prefactor underestimates the prefactor value while the Tait prefactor overestimates the prefactor value.
Indeed, this conclusion agrees with the work of \citet{Pantaleone2025}, which stated that the experimental peak lays in between the two peaks computed using Tait and Campbell approaches.

\subsection{Comparison with theoretical studies}

\begin{table*}[h!]
\centering
\caption{Values of the CO$_2$ BE as computed in different theoretical studies using various methods and ice grain models. Units are in K. Max and Min are maximum and minimum values, $\mu$ is the distribution’s mean (or means).}
\begin{tabular*}{\linewidth}{@{\extracolsep{\fill}} l c c c c @{}}
\hline
\hline
BE (K) & Surface Model & Methodology & BSSE/ZPE & Ref. \\
\hline
3100 & Single water molecule & M06-2X/aug-cc-pVTZ & No/No & \tablefootmark{a} \\ \hline

1506 (monomer) & \multirow{5}{*}{Water clusters} & \multirow{5}{*}{MP2/A-VDZ}  & \multirow{5}{*}{No/No} & \multirow{5}{*}{ \tablefootmark{b}} \\ 
1935 (trimer)  &  &  &  &  \\ 
2293 (tetramer)&  &  &  &  \\ 
2287 (pentamer)&  &  &  &  \\ 
2352 (hexamer) &  &  &  &  \\ 
\hline
min: 1104 & \multirow{2}{*}{AWS cluster model 480 H$_2$O}  & \multirow{2}{*}{AKMC simulations} & \multirow{2}{*}{---} &  \multirow{2}{*}{\tablefootmark{c}} \\ 
max: 3240 &  &  &  &  \\ \hline
min: 1489 & \multirow{2}{*}{Periodic AWS 60 H$_2$O} & \multirow{2}{*}{B3LYP-D3/A-VTZ//HF-3c} & \multirow{2}{*}{Yes/Yes \tablefootmark{*}}  & \multirow{2}{*}{\tablefootmark{d}}\\
max: 2948 & & & & \\ \hline
$\mu_1$:1408 & \multirow{3}{*}{AWS clusters 22 H$_2$O} & \multirow{3}{*}{$\omega$–PBE/def2-TZVP//HF-3c} &  \multirow{3}{*}{Yes/Yes} & \multirow{3}{*}{\tablefootmark{e}} \\ 
$\mu_2$:1819 & & & & \\ 
max: 2389& & & & \\ \hline
min: 1595 & \multirow{2}{*}{Periodic AWS 60 H$_2$O} & \multirow{2}{*}{B3LYP-D3/A-VTZ} & \multirow{2}{*}{Yes/Yes} & \multirow{2}{*}{\tablefootmark{f}} \\
max: 3135 & & & & \\ \hline
min: 1158 & \multirow{4}{*}{Iced grain 200 H$_2$O} &\multirow{2}{*} {ON(DLPNO-CCSD(T)/ }& \multirow{4}{*}{Yes/Yes} & \multirow{4}{*}{This work} \\ 
$\mu_1$:1649 & & & & \\
$\mu_2$:2339 & & \multirow{2}{*}{aug-cc-pvtz:GFN2)//ON(B97-3C:GFN2)} & & \\
max:3049 & & & & \\ \hline
 \end{tabular*}
\label{tab: computational BEs}
\tablefoot{
\tablefoottext{*}{ZPE introduced in the form of a scaling factor.}
\tablefoottext{a}{\citet{Wakelam2017}}
\tablefoottext{b}{\citet{Das_2018}}
\tablefoottext{c}{\citet{Kerssemeijer2014_co2}}
\tablefoottext{d}{\citet{Ferrero2020}}
\tablefoottext{e}{\citet{Bovolenta2022}}
\tablefoottext{f}{\citet{Bulik2025}}
\
}
\end{table*}

This section is dedicated to the comparison of our results to previous theoretical works, including the discussion on the choice of the surface model and its parameters, of the methodology and its consequences, and the improvements provided by this work compared to values already available in literature (presented in Table \ref{tab: computational BEs}).
The first work to ever compute BEs using theoretical chemistry methods was \citet{Wakelam2017}. They used a simple surface model, where the ice is represented by a single water molecule, which, combined with the lack of ZPE and BSSE corrections resulted in high values of the CO$_2$ BE. 
Although the computational methodologies in this work were limited, it inspired the next works by different groups. 
The following study conducted by \citet{Das_2018} extended the idea of \citeauthor{Wakelam2017} by calculating the BE of CO$_2$ adsorbed on water clusters, ranging from a single water molecule to six water molecules. 
For their methodology, they used an accurate wave function-based perturbation theory method, although with a limited basis set.
Their values (1506--2352 K) correspond with a part of the higher BE region of the distribution presented in this work. 
However, they did not consider the limitations of the surface model and the BSSE and ZPE corrections.
\citet{Kerssemeijer2014_co2} subsequently used \textit{ab initio} Kinetic Monte Carlo methods to simulate the adsorption of CO$_2$ clusters on an AWS model composed of 480 water molecules. 
The cited values are reported at 50 K. 
The size of their surface model assures that all BE sites on their study were explored. 
Although \citet{Kerssemeijer2014_co2} do not report the statistical parameters of their distribution, the BE range established by their work corresponds well with the range found in this study.
This serves as a confirmation that the presented data fully samples all binding sites on the AWS ice model.
\begin{figure}[]
    \centering
    \includegraphics[width=1\linewidth]{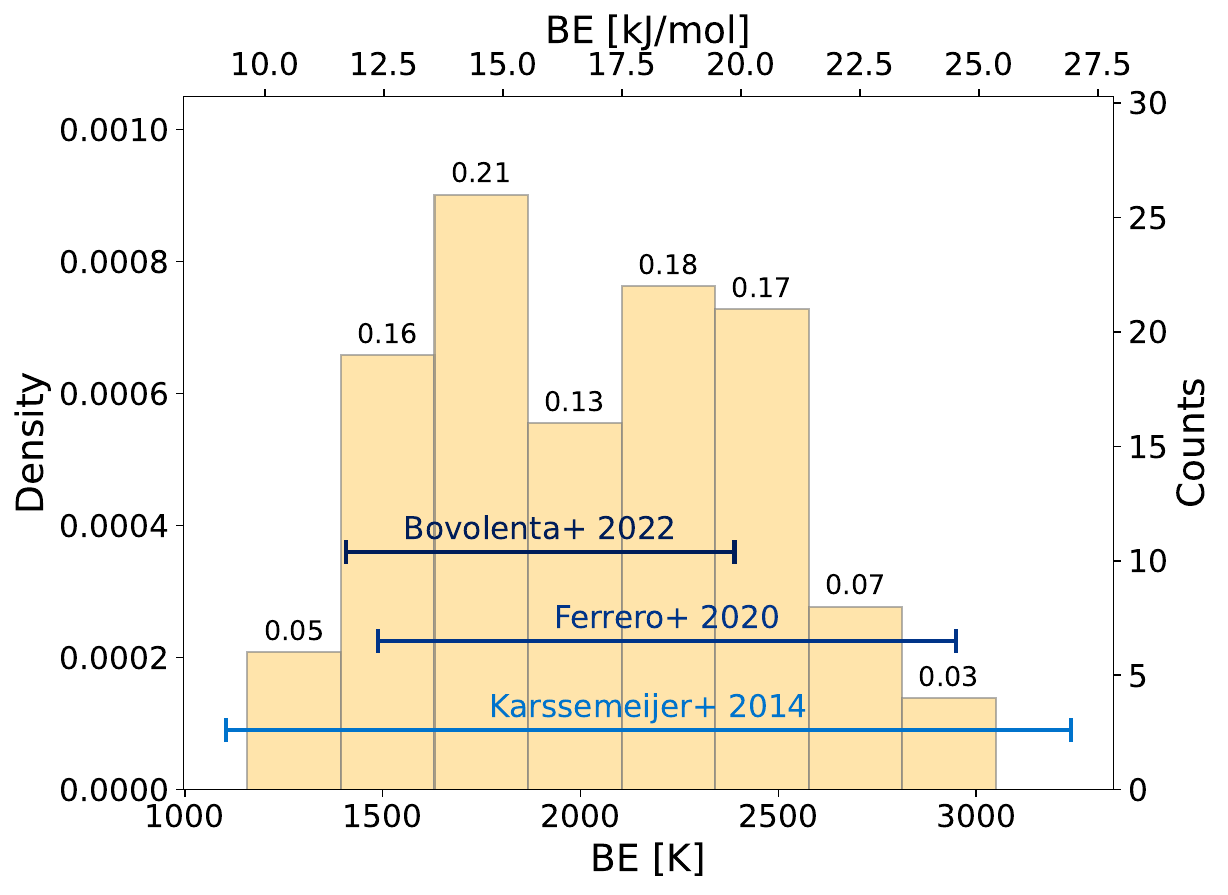}
    \caption{The BE distribution of this work with BE ranges from the three most relevant studies}
    \label{fig:Be_ranges}
\end{figure}
\citet{Ferrero2020} was the first work to compute BEs on periodic AWS models.
In their study, the ONIOM method by using a cost-effective HF-3c method and then refining the energy using the B3LYP-D3 method was adopted. 
In this work, the ZPE correction was introduced in the form of a scaling factor, based on calculations on crystalline ices, which is later adopted for amorphous ice surfaces.
This approach was subsequently used in \citet{Perrero2022,Martínez-Bachs_2024} for S- and N-bearing molecules, and in \citet{Kakkar_2025} for iCOMs.
The BEs obtained by \citet{Ferrero2020} were verified in a work done by some of us \citep{Bulik2025}, where the geometries were fully relaxed at B3LYP-D3 level and the ZPE was calculated for each binding site. 
The BE ranges obtained at full DFT were slightly higher than the original ones but still consistent with the previous results of \citet{Ferrero2020}. 
This could be partially explained by the fact that B3LYP-D3 is known to overbind species exhibiting H-bonding and HF-3c could identify different PES minima than B3LYP-D3.
The study by \citet{Bovolenta2022} employs a similar approach as the one of \citet{Ferrero2020}, where the geometries are obtained with HF-3c and then refined at DFT level. 
In this case, a different functional and basis set were selected and the surface was simulated using different clusters of 22 water molecules.
The result was a BE distribution exhibiting a faint bi-modality. 
The range of the distribution covers most of the BEs found in this study with an exception of the more extreme values, which explains the discrepancy in their distribution parameters and the ones described in this work.

\subsection{Effect of the adsorption on the spectral features}

\begin{figure}
    \centering
    \includegraphics[width=1.0\linewidth]{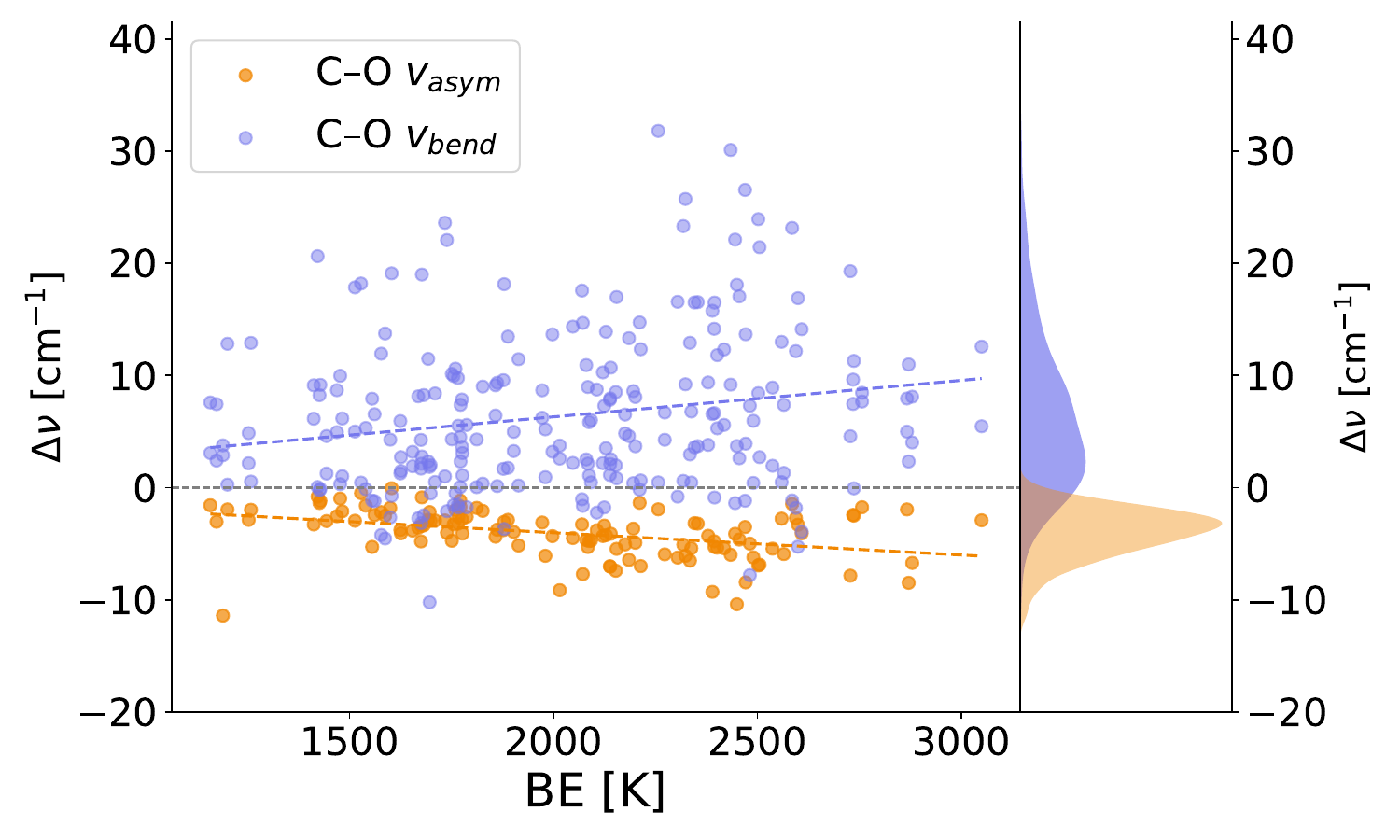}
    \caption{Plot showing the frequency shift $\Delta\nu$ for the asymmetric stretching mode (purple) and the bending mode (orange). On the right side, KDE plot for the $\Delta\nu$ for each mode is displayed. }
    \label{fig:freq_shift}
\end{figure}

Adsorption can affect the peak position of each vibrational mode in the infrared spectrum with respect to the non-adsorbed state. 
This effect can be quantified by the difference between gas-phase frequency and the one of the adsorbed species, defined as $\Delta\nu = \nu_{gas} - \nu_{ads}$. 
These values for a single, most stable, binding site were reported for the most relevant icy species by some of us in a previous work \citep{Bulik2025}. 
In this work, we compute the $\Delta\nu$ for all binding sites and show its correlations with the BE along with the kernel density estimate (KDE) for each IR active mode of CO$_2$ (Figure \ref{fig:freq_shift}).
The KDE represents the shape of the absorption bands without the contribution of the broadening effects of pressure or temperature. Nevertheless, the KDE recovers the shapes that can be observed in the spectra taken in the experiments or against the radiation of the background star (in the case of molecular clouds) or the central star (in the case of protoplanetary disk). 

In the case of the bending mode (orange colour) the peak becomes shifted to the higher frequency compared with the gas phase species.
While for the asymmetrical stretching (violet colour) there is a shift towards lower frequencies with respect to the gas phase. 
This is rooted in the geometry of the vibrational mode and the way CO$_2$ bonds with the surface.
The ipsochromic vibrational shift of the CO$_2$ bending mode is expected as the interaction with ice renders the motion more difficult. On the contrary, the H-bond interaction between the oxygen of CO$_2$ and ice weaken the force constant of the C=O bond causing the bathochromic shift of the asymmetric stretching frequency.

The analysis of the scatter plot reveals that the two active normal modes (i.e., asymmetric stretching and bending) do not exhibit a meaningful correlation with the BE. For both modes, the larger the BE, the higher the absolute value of $\Delta\nu$. 
The stretching mode shows less divergence, with the width of the distribution being less than 12 cm$^{-1}$, with the mean being $-4.1$ cm$^{-1}$ and the extreme values being $-11.4$ and $-0.1$ cm$^{-1}$, while for the bending mode the distribution width is more than 40 cm$^{-1}$ with a mean of 6.3 cm$^{-1}$, with the extreme values being $-10.2$ ad 31.8 cm$^{-1}$.
For the stretching mode, the adsorption causes a blueshift of the peak position, and for the bending mode it causes both a redshift and a blueshift.

\subsection{Astrophysical implications}

\begin{figure*}
    \centering
    \includegraphics[width=0.85\linewidth]{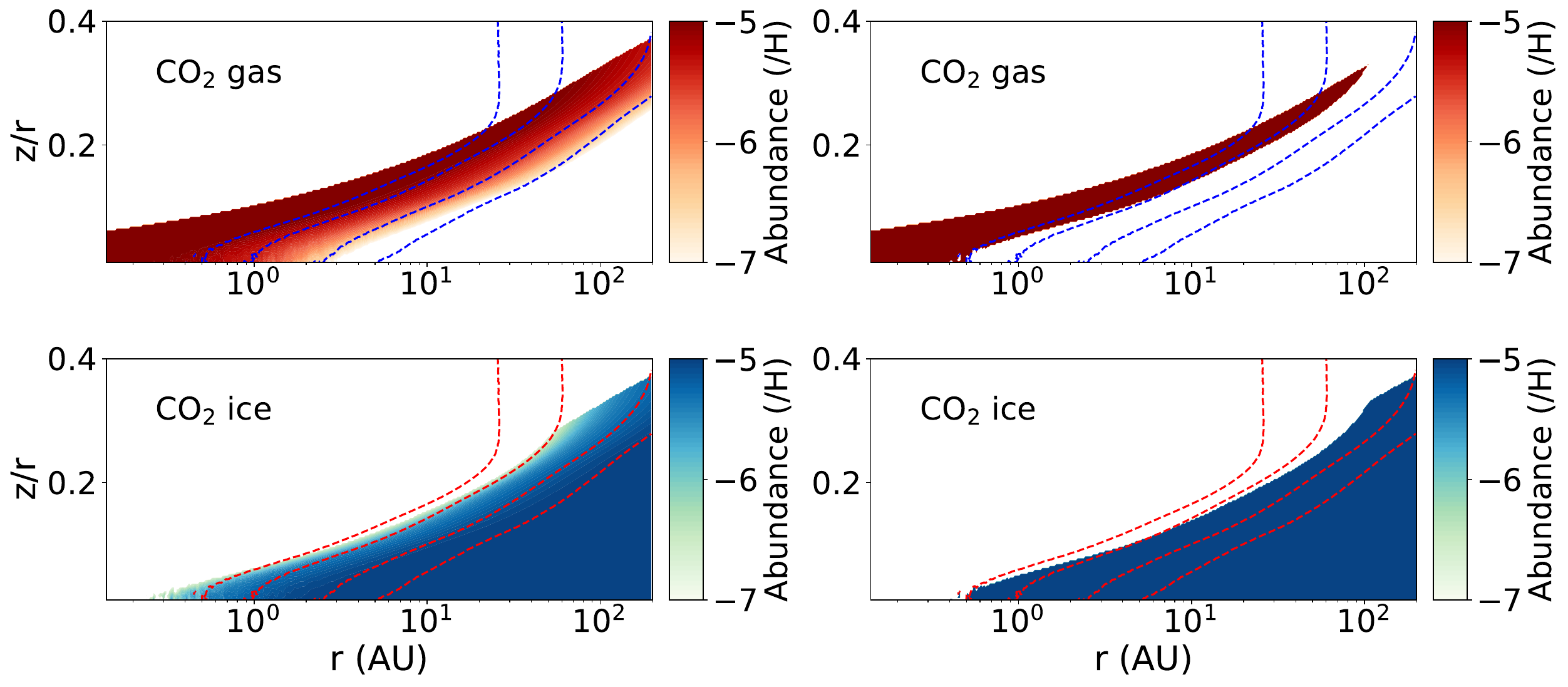}
    \caption{Two dimensional spatial distributions of CO$_2$ gas (upper panels) and ice (lower panels) abundances on the disk, shown using a multibinding description (left panels) and a single binding description (right panels). The vertical axes represent height normalized by radius. Dashed lines depict the positions where the dust temperature is equal to 100, 80, 60, 40 and 30 K.}
    \label{fig:ppd_distribution_vs_single}
\end{figure*}

\begin{figure*}
    \centering
    \includegraphics[width=0.80\linewidth]{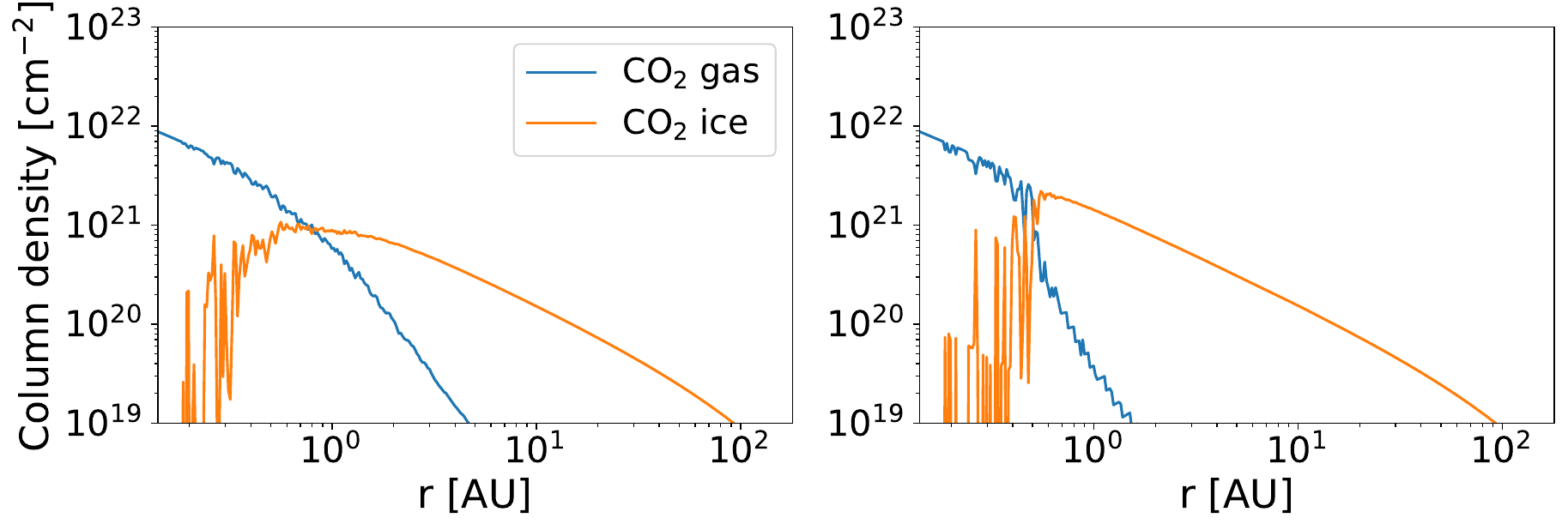}
    \caption{Vertically integrated CO$_2$ gas and ice column densities as functions of radius. The left panel shows the model with the multibinding
description, while the right panel shows the single binding description.}
    \label{fig:co2_column}
\end{figure*}

\begin{figure*}
    \centering
    \includegraphics[width=0.85\linewidth]{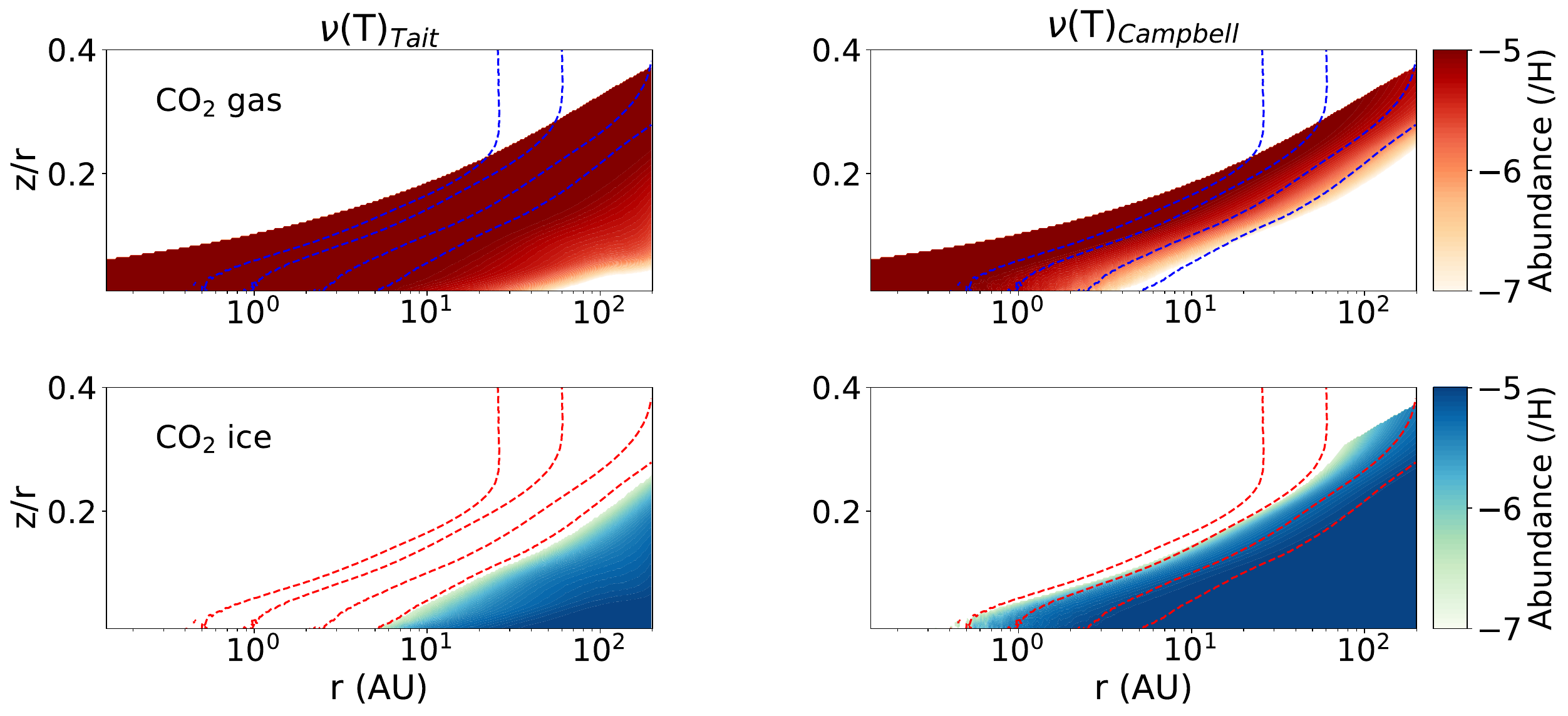}
    \caption{Two dimensional spatial distributions of CO$_2$ gas (upper panels) and ice (lower panels) abundances on the disk, shown using a multibinding description and different approaches to model the prefactor $\nu$(T)$_{Campbell}$ (right), $\nu$(T)$_{Tait}$ (left). The vertical axes represent height normalized by radius. Dashed lines depict the positions where the dust temperature is equal to 100, 80, 60, and 30K.}
    \label{fig:ppd_distrib_prefactors}
\end{figure*}

\begin{figure}
    \centering
    \includegraphics[width=0.99\linewidth]{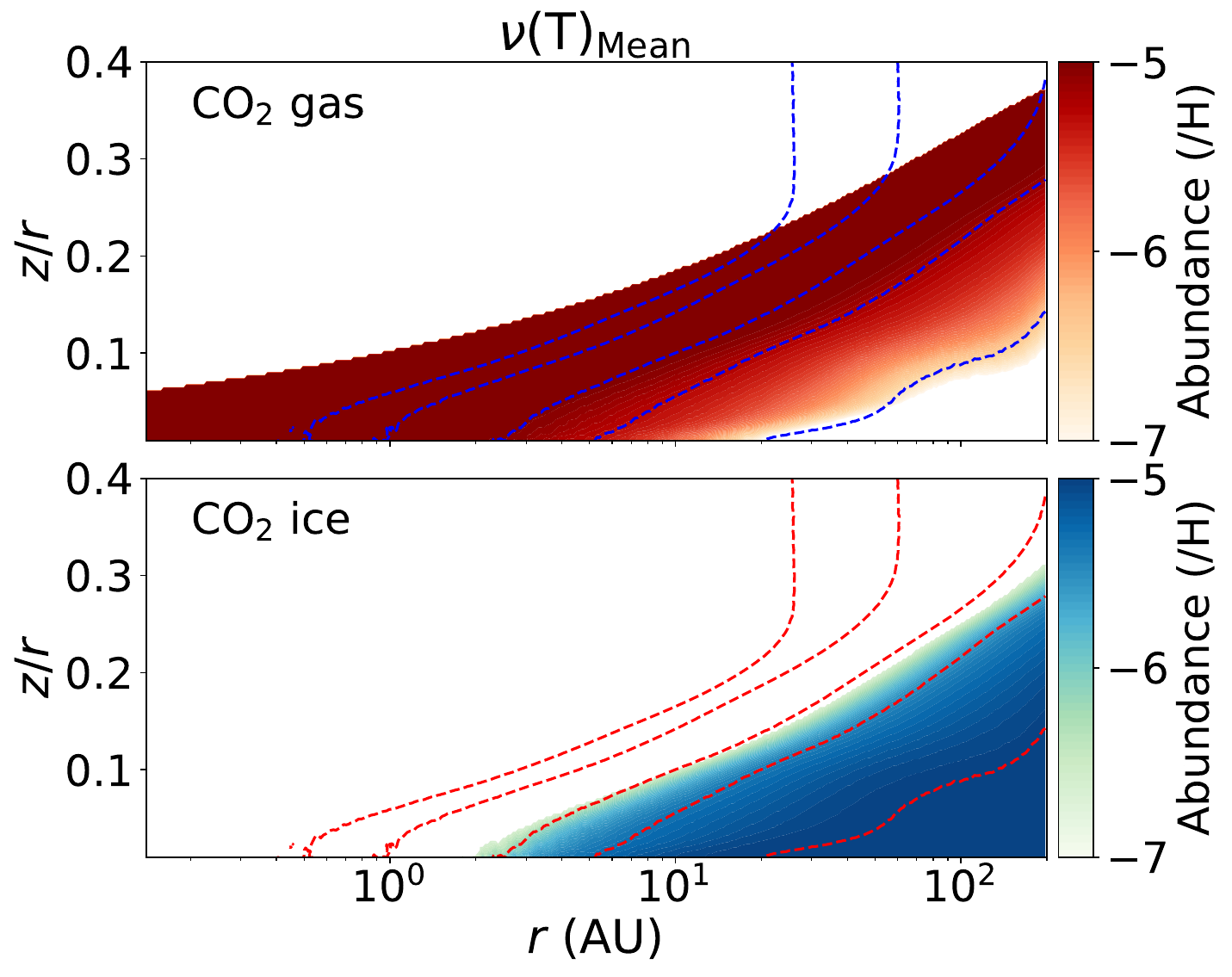}
    \caption{Two dimensional spatial distributions of CO$_2$ gas (upper panels) and ice (lower panels) abundances on the disk, shown using a multibinding description and the geometrical mean between the prefactor $\nu$(T)$_{Campbell}$ and $\nu$(T)$_{Tait}$. The vertical axes represent height normalized by radius. Dashed lines depict the positions where the dust temperature is equal to 100, 80, 60, 30 and 18 K.}
    \label{fig:geom_mean}
\end{figure}


The distribution of $\text{CO}_2$ between its ice and gas phases within a protoplanetary disk (PPD) is one of the carriers that impacts the local C/O elemental ratio \citep{Turrini_2021}.
This ratio, in turn, dictates the molecular and elemental composition of planets forming within that disk region. 
Furthermore, because $\text{CO}_2$ is a common molecule found in exoplanetary atmospheres, studying its evolution and phase distribution in the PPD provides critical context regarding the molecular inheritance passed down from the disk to the final planetary system.
This property can be inferred from isotope ratios, using the $^{13}$CO$_2$/$^{12}$CO$_2$ of either gas or ice observations \citep{Brunken2024, Brunken2024a}.

In this work, we aim to evaluate what is the impact of introducing our predicted CO$_2$ BE distribution on its gas and ice abundances and distributions in a PPD.
Previous studies already discuss this effect for water \citep{Tinacci2023} and CO \citep{Bovolenta2025}. 
However, these works report single Gaussian BE distributions, while here we report a bimodal distribution with almost equal partitioning of the two Gaussians.
We adopted the procedure described by \citet{Bovolenta2025} to calculate the partitioning between the gas and solid phases at each position in a 2D protoplanetary disk model for CO$_2$.
For the disk model, we adopted the density and temperature profiles of dust and gas along with the UV radiation field relevant for the TW Hya disk \citep{Cleeves_2015, Furuya2022}.
The total (gas + solid) CO$_2$ abundance throughout the disk assumed for this case was 10$^{-5}$ relative to the H nuclei. 
To account for the dissociation of CO$_2$ under the strong UV radiation coming from the central star we assume that no CO$_2$ molecules exist in regions with A$_{V}$ < 1 mag \citep{Aikawa_2006}, similarly to assumptions in \citet{Bovolenta2025}.

Two ways to calculate the partitioning of CO$_2$ between the gas and the solid phase were employed. In the first method (multibinding approach), we used the BE distribution to calculate the partitioning at each disk position as in \citet{Tinacci2023, Bovolenta2025}.
We calculated the fraction of adsorption sites where $k_{\mathrm{thdes}}$ >
$k_{\mathrm{ads}}$ ($f_{\mathrm{gas}}$), and the gas-phase and solid-phase CO$_2$ abundances at each disk position were given by $10^{-5} f_{\mathrm{gas}}$ and $10^{-5} \left(1 - f_{\mathrm{gas}}\right)$, respectively, where, $k_{\mathrm{thdes}}$ and $k_{\mathrm{ads}}$ are rate constants for thermal desorption (see Equation \ref{eq:desorption}) and adsorption, respectively.
In the second approach (single binding approach), we use the mean BE value of the second peak from our BE distribution (BE = 2339 K) to estimate the $k_{\mathrm{thdes}}$, which is compared with $k_{\mathrm{ads}}$ at each disk position. All CO$_2$ molecules were assumed to exist only in vapour when $k_{\mathrm{thdes}}$ > $k_{\mathrm{ads}}$, while for the opposite relation all CO$_2$ abundance was assumed to be in ice form.
In the first analysis we assumed the prefactor ($\nu$) to be equal to 10$^{12}$ s$^{-1}$, as in \citet{Bovolenta2025}.
The results of these simulations are shown in Figure \ref{fig:ppd_distribution_vs_single}, which shows the 2D (radial + vertical) distribution of gas-phase and solid-phase CO$_2$ using the multibinding description (right) and the single-binding description (left).
In the single binding approach, there is a clear border between the ice and gas phase, with the CO$_2$ snowline forming around 80 K in the upper parts of the disk and then at 100 K in the mid-plane.
Meanwhile, in the multibinding approach, the snowline becomes more diffused and CO$_2$, while it is formed in the same regions of the disk, the gas fraction is much more extended and can be present in regions of 30--40K.
We envision no clear additional effect of having a bimodal Gaussian distribution of the BEs.
Consistently with previous studies, we find that, in the multibinding approach, the region of the disk where gas and ice exist simultaneously is extended compared to the single binding approach.
This is also evident in the vertically integrated CO$_2$ column densities shown in Figure \ref{fig:co2_column}.

The effect of the prefactor on the gas/ice partitioning was also explored \citep[see, e.g.,][]{Ceccarelli2017}.
In the previous section, we showed two models to calculate the prefactor and use them to compare the experimental results with our theoretical findings, see Section \ref{experiment}.
The temperature dependent formula for the two prefactors, $\nu(T)_{\mathrm{Tait}}$ and $\nu(T)_{\mathrm{Campbell}}$, was incorporated into the code, to calculate the specific prefactor at each point of the disk and then obtain the desorption time as described above and compare it with the adsorption time.
This was done in the multibinding approach for both types of prefactor.
The result is shown in Figure \ref{fig:ppd_distrib_prefactors}.
When $\nu(T)_{\mathrm{Campbell}}$ is used for the calculation of the gas/ice partitioning in the disk (on the right), the CO$_2$ snowline is formed around 80--100K similarly to when a constant prefactor ($\nu$=10$^{12}$ s$^{-1}$) is used (see the left panel of Figure \ref{fig:ppd_distribution_vs_single}).
For the gas-phase CO$_2$, the model including the Campbell prefactor is slightly more extended to a 60K grain temperature in the mid-plane, while when the constant prefactor is used, gaseous CO$_2$ extends only to where the grain temperature is 80K.
However, when $\nu(T)_{\mathrm{Tait}}$ is used in the simulation along with the multibinding approach, there is a significant shift of the position of the snowline formation. 
In this case, the CO$_2$ snowline is formed in the region where the grain temperature is below 30K (see Figure \ref{fig:ppd_distrib_prefactors} left), which is 50K lower than the values reported for the constant prefactor (10$^{-12}$ s$^{-1}$) and $\nu(T)_{\mathrm{Campbell}}$.
Furthermore, the gas fraction of CO$_2$ is further extended and can exist in regions where the grain temperature is below 15K.
The Tait prefactor, along with the BE distribution was also used by \citet{Tinacci2023} to show the partitioning of water in the disk.
They also report a significant difference between that approach and the single binding and constant prefactor approach.
In this work, we attempted to compare the two prefactor models.
However, the comparison of the theoretical TPD simulations with the experimental spectra did not indicate which prefactor could be more reliable (see Section \ref{experiment}). 
This is consistent with the conclusions of \citet{Pantaleone2025}, suggesting that a suitable prefactor lays between the two proposed models.
Therefore, we sought to use the geometrical mean between the two values for each temperature.
We calculated the fractionation using the resulting mean prefactor at each point of the disk (Figure \ref{fig:geom_mean}).
In this case, the CO$_2$ snowline is formed at 60K and the gas fraction extends until 18K.
This maybe crucial, since the observations in the near infrared of the CO$_2$ molecular line probe the inner disc, while our simulations show that the gas phase can indeed extend further into the outer disk. 

\section{Conclusions}

In this work, we provide a comprehensive study of the CO$_2$ BEs and spectral properties, computed on an iced grain model composed of 200 water molecules representative of an interstellar ice mantle covering a dust grain.
The ice model was constructed using the ACO-FROST procedure \citep{Germain2022}. 
The procedure to calculate the BEs was adopted based on the previous works within the group \citep{Germain2022,Tinacci2023, Bariosco2024, Bariosco2025}.
The main results are:
\begin{itemize}
    \item Following our procedure, a BE distribution was obtained from the 122 unique binding sites identified on the grain. 
    The distribution follows a bimodal Gaussian function, whose parameters are obtained from a bootstrapping method.
    The lower peak has $\mu_1$=1648K with $\sigma_1$=229K, while the higher peak $\mu_2$=2339K with $\sigma_2$=274K. The relative intensity of each peak is w$_1$=0.46 and w$_2$=0.54.
    \item We propose a new multi-molecule adsorption scheme, to investigate the intermolecular interactions of the adsorbate on the iced grain.
    The results show that there is little bonding co-operativity when multiple CO$_2$ are adsorbed in the "far" regime, being dispersed on the surface.
    However, when a cluster of CO$_2$ molecules is created on top of the iced grain two trends appear: when the CO$_2$ cluster consists of up to 6 molecules, the BEs fall in the second (higher) peak of the BE distribution, while when more molecules are subsequently adsorbed the BEs become lower and fall in the first (lower) peak.
    \item Calculated BE values were compared with the available experimental values. 
    This was done by adopting the methodology to model the surface coverage based on the work of \citet{Bovolenta2025}.
    This work employs two prefactor models ($\nu_{\mathrm{Tait}}$ and $\nu_{\mathrm{Campbell}}$) to simulate the TPD experimental spectra.
    The values are consistent with the experimental results; however, the experimental values tend to favour the higher end of the BE distribution. The comparison showed that the prefactor models, either underestimate (Campbell) or overestimate (Tait) the prefactor value.
    \item The comparison with the theoretical values shows that the values presented in this work are consistent with the previously reported values.
    Moreover, the comparison with studies that report BE distributions (or distribution-like results), confirms that in this work the AWS is fully sampled by out methodology.
    \item We showed the impact of adsorption of the CO$_2$ molecule on an AWS on the spectral features by plotting the computed BEs with the computed divergence of the central peak of the adsorbate from the gas phase ($\Delta\nu$). 
    Although there is no strong correlation between the frequencies and the BE, the KDE plot of the $\Delta\nu$ recovers the broadening of the peaks of both vibrational modes detectable in star-forming regions.
    \item We find that employing the BE distribution to model the ice/gas partitioning of CO$_2$ has a small effect on the snowline formation; however, our simulations show that it extends the region where gas and ice phases are present in the disk. 
    This is consistent with the previous result of \citet{Bovolenta2025} for CO. 
    \item We find there is a significant impact of the prefactor on the gas/ice partitioning in the disk.
    For the first time, we compare the three different prefactor models to calculate the gas/ice partitioning in a PPD.
    We use one constant prefactor ($\nu$ = 10$^{12}$ s$^{-1}$) and two temperature-dependent prefactors to simulate the desorption timescale ($\nu(T)_{\mathrm{Tait}}$ and $\nu(T)_{\mathrm{Campbell}}$).
    The result shows little difference between the constant prefactor and $\nu(T)_{\mathrm{Campbell}}$.
    However, when $\nu(T)_{\mathrm{Tait}}$ is employed in the PPD simulation, the snowline is formed at a temperature 50K lower than in the previous simulations. This is tackled with a straightforward geometrical mean between the two prefactors, which was calculated at each point of the disk. 
    The final snowline is formed at 60K and the gas fraction is extended until 18K. Thus showing that the gas and ice phases coexist in an extended region of the disk.
\end{itemize}
Overall, this study highlights the necessity of moving beyond single-value binding energies by demonstrating how BE distributions and temperature-dependent prefactors significantly shift the predicted CO$_2$ snowline and gas-ice partitioning. 
By reconciling our computational results with experimental TPD spectra and identifying the influence of the $\nu(T){\mathrm{Tait}}$ and $\nu(T){\mathrm{Campbell}}$ prefactors, we provide a more physically grounded framework for modelling the chemical evolution of PPDs. 

\begin{acknowledgements}
      Part of this work was supported by funding from the European Union's Horizon 2020 research and innovation program from the European Research Council (ERC) for the project “Quantum Chemistry on Interstellar Grains” (QUANTUMGRAIN), grant agreement no 865657. MICIN (projects PID2024-157971NB-C21 and CNS2023-144902) is also acknowledged. The Italian Space Agency for co-funding the Life in Space Project (ASI N. 2019-3-U.O), the Italian MUR (PRIN 2020, Astrochemistry beyond the second period elements, Prot. 2020AFB3FX) are also acknowledged for financial support. Support from the Project CH4.0 under the MUR program “Dipartimenti di Eccellenza 2023-2027” (CUP: D13C22003520001) is also acknowledged.  The authors thankfully acknowledge the supercomputational facilities provided by CSUC and MareNostrum5. AR acknowledges Accademia delle Scienze di Torino for supporting the project “In silico interstellar grain-surface chemistry” and gratefully acknowledges support through the 2023 ICREA Award.
\end{acknowledgements}

\bibliographystyle{bibtex/aa}          
\bibliography{bibtex/co2_bib}

\newpage
\newpage

\begin{appendix}

\section{Pruning procedure}
\label{pruing}

The pruning procedure was developed to ensure a uniform sampling of BE on the AWS, since during the geometry optimization, some structures may converge to the similar minima in the potential energy surface (PES).
This is caused by the mild nature of the interaction, PES complexity and number of explored binding sites.
In order to ensure that the binding sites are sampled uniformly we pruned the redundant structures using the procedure described in \citet{Bariosco2024}.
The pruning was performed twice, first on the geometries obtained at XTB-GFN2 level and then again after the DFT optimization at B97-3c.
This is because the benchmark showed that XTB-GFN2 finds two minima for the water carbon dioxide dimer while the DFT-based method found only one.
The pruning criteria were selected based on the correlation between the RMSD and the energy difference between the two PES minima ($\Delta E_{\mathrm{C_{ji}}}$).
\begin{figure}[h!]
    \centering
    \includegraphics[width=0.95\linewidth]{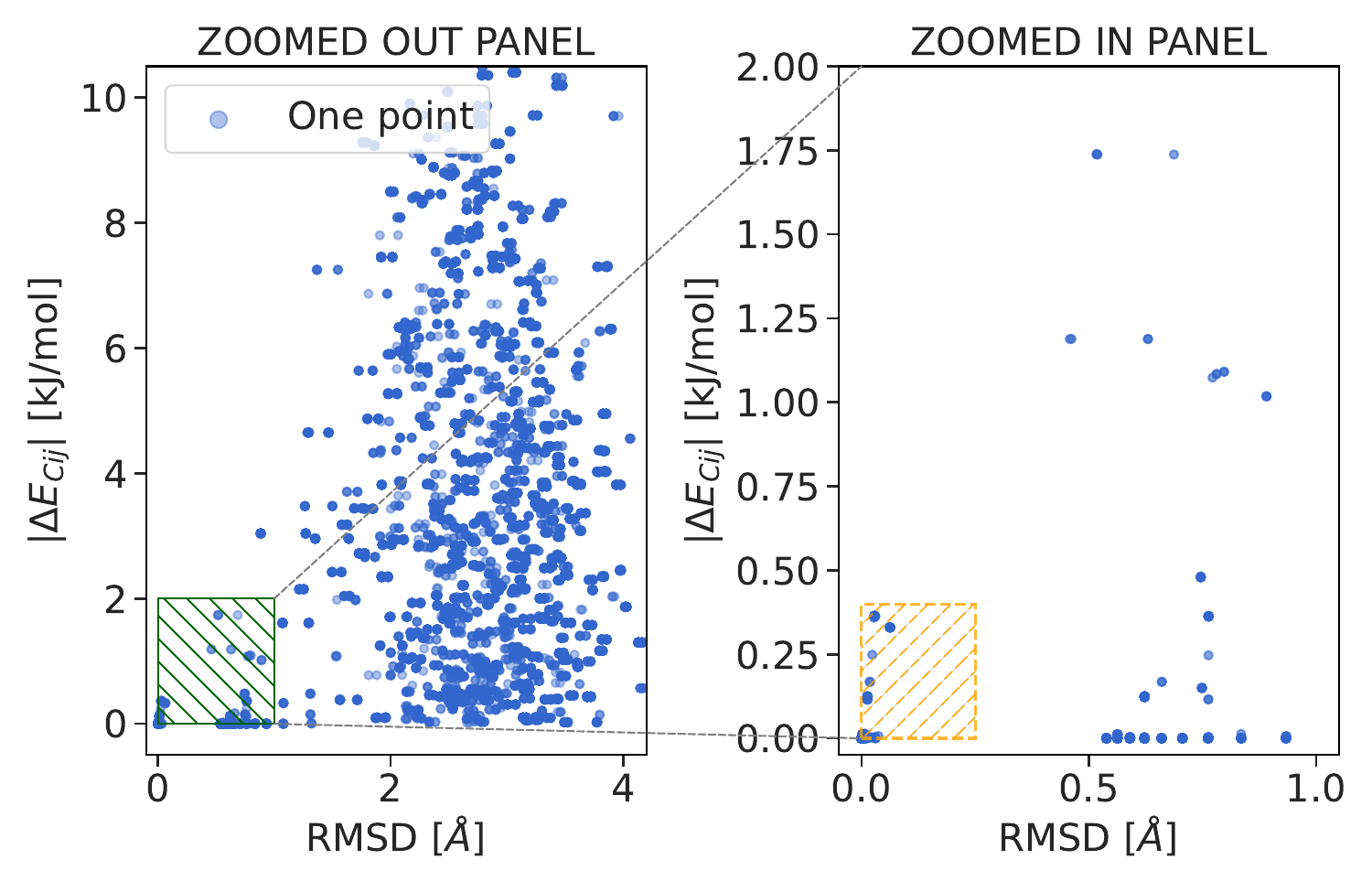}\\[1em]
    \includegraphics[width=0.95\linewidth]{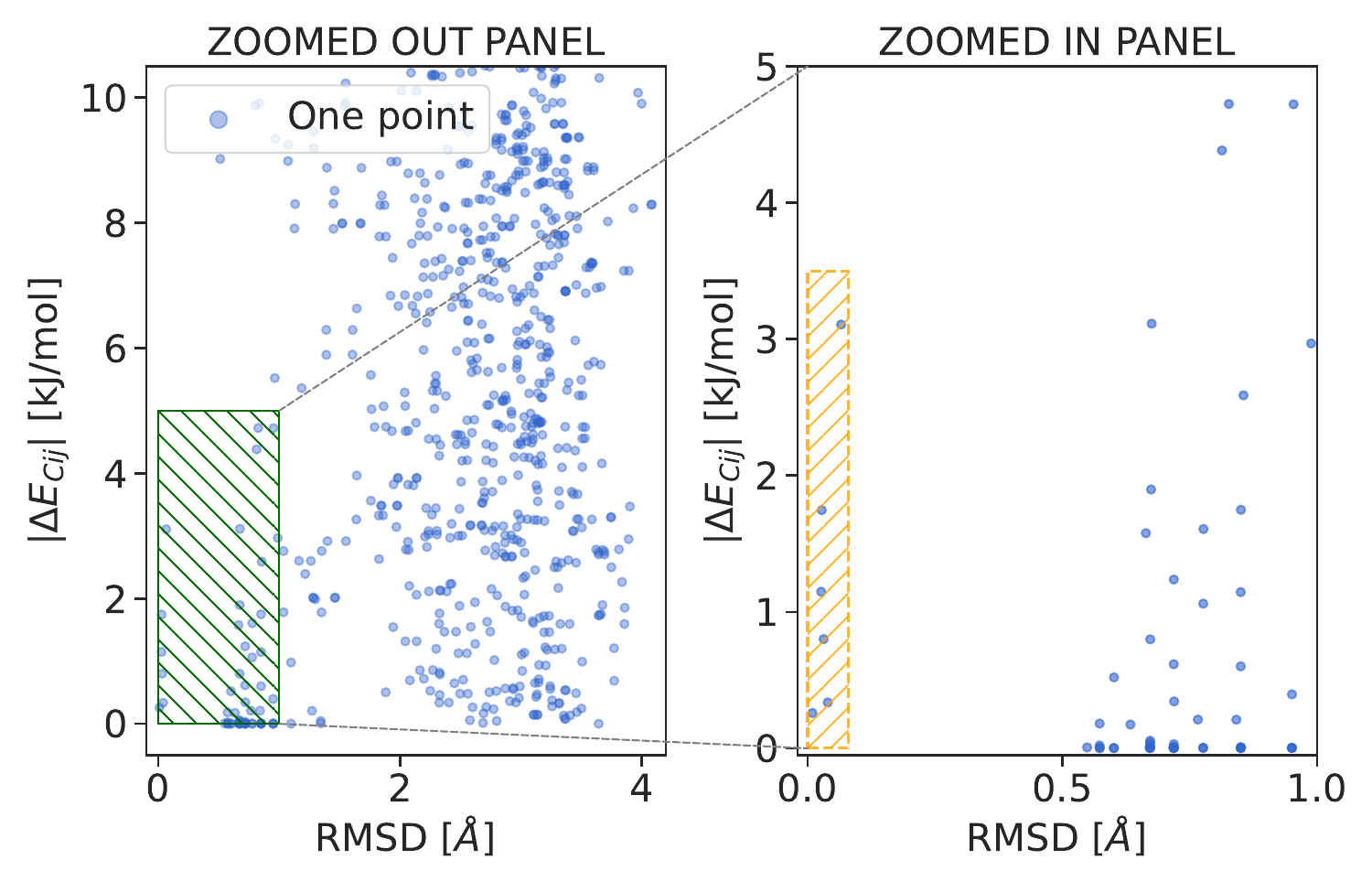}
    \caption{The correlation between the RMSD and $\Delta E_{\mathrm{C_{ji}}}$, the top plot shows the pruning at XTB-GFN2 level and the bottom plot shows the prunnig at DFT level. The left side is the total scatter plots are reported and the right side is the zoomed in part that is marked as the green rectangle in the left panels. The orange rectangle marks the criteria where the redundant structures were identified.}
    \label{fig:pruning}
\end{figure}
The scatter plots showing this relationship for both levels of theory are shown in Figure \ref{fig:pruning}.
The pruning at XTB-GFN2 level (upper plots) was run on 486 structures and in the procedure identified 312 redundant geometries leaving 174. 
The geometry of the remaining structures as refined at DFT level using the ONIOM scheme (see Section \ref{method}) and then the pruning procedure was performed again (lower plot).
This resulted in the identification of further 52 redundant geometries, which leaves the final 122 unique binding sites.

\section{Benchmark details}

\subsection{DLPNO-CCSD(T) vs B97-3c vs GFN2}

\begin{table*}[h!]
    \centering
    \begin{tabular*}{\linewidth}{@{\extracolsep{\fill}} l c c c r  @{}} \hline \hline
    \multicolumn{5}{c}{$\text{H}_2\text{O}\cdots\text{CO}_2$} \\ \hline
    Level of theory  & BE*(kJ/mol) & BE (kJ/mol) & H$_2$O $\cdots$ CO$_2$ & O$\mathrm{\hat{H}}$C\\
    CCSD(T) aug-cc-pVTZ aug-cc-pVTZ/C & 12.31 & 8.79 & 2.75 & 127.5 \\
    CCSD(T)/aug-cc-pvtz //B97-3c & 11.03 & 8.93 & 2.92 & 126.65 \\
    DLPNO-CCSD(T)/aug-cc-pvtz //B97-3c & 10.54 & 7.84 & 2.92 & 126.65 \\
    DLPNO-CCSD(T)/aug-cc-pvtz //$\omega$B97x-3c & 11.12 & 7.70 & 2.80 & 125.24 \\
    B97-3c & 11.31 & 9.16 & 2.92 & 126.65 \\
    $\omega$B97x-3c & 13.39 & 9.98 & 2.80 & 125.24 \\
    HF-3c & 16.88 & 12.69 & 2.59 & 125.79 \\ 
    GFN2 & 10.21 (7.90) & --- & 2.66 (2.15) & 125.21 (134) \\ \hline 
    \end{tabular*}
    \caption{The level of of theory and according BE* (no ZPE) and the BE (fully corrected for BSSE and ZPE) are reported in kJ/mol, for the dimer (top part of the table) the bond length (between the O atom from H$_2$O) and the C atom) and the bond angle are reported. For GFN2 we report the energy of both minima identified by the method, the one not found by DFT is shown in parenthesis.}
    \label{tab:benchmark}
\end{table*}

\begin{table}[h!]
    \centering
    \begin{tabular*}{\linewidth}{@{\extracolsep{\fill}} l r @{}} 
    \hline \hline
    \multicolumn{2}{c}{Bigger system: CO$_2$ on the grain} \\ \hline
    Level of theory  & BE* (kJ/mol) \\ \hline

    ON(DLPNO-CCSD(T)/aug-cc-pvtz:GFN2)// & \multirow{2}{*}{13.62} \\
    ON(B97-3c:GFN2) & \\ \hline

    ON(DLPNO-CCSD(T)/aug-cc-pvtz:GFN2)// & \multirow{2}{*}{13.97} \\
    ON($\omega$B97x-3c:GFN2) & \\ \hline 
    
    \end{tabular*}
    \caption{Benchmark details for the bigger system only the uncorrected BE* (no BSSE, no ZPE correction) is given in kJ/mol.}
    \label{tab:bigger_bench}
\end{table}

A thorough benchmark was done to ensure the accuracy of the computed BE values.
The benchmark details are shown in Table \ref{tab:benchmark}. 
The energies and geometry parameters (bond length and angle) are reported.
During the benchmark, the geometry and the energy of the H$_2$O$\cdots$CO$_2$ dimer was investigated.
The parameters were compared against the computational golden standard CCSD(T). 
The result showed, that at B97-3c level the bond length of the dimer was longer than at CCSD(T),
while $\omega$B97x-3c showed quite a good match.
For the XTB-GFN2 (GFN2), this lower accuracy method identified two (PES) minima, the first one like the one at CCSD(T), where CO$_2$ interacts through the carbon atom with the oxygen in the water molecule. 
The second one (its parameters are given in the parenthesis), where CO$_2$ interacts through the oxygen atom with one of the hydrogens in the water molecule, forming a weak hydrogen-bonded species.
This was the reason, why the pruning procedure was repeated after the geometry refinement (see Section \ref{pruing}).
When looking at the energies, the a single point at DLPNO-CCSD(T)/aug-cc-pvtz for the geometries obtained with both of the DFT-based methodologies was around 1 kJ/mol lower than the full CCSD(T), which is within the computational error.
To ensure that the lower end of the BE distribution is not affected by the longer bond length identified at B97-3c, a benchmark on the bigger system was performed.
One of the lower BE sites was selected and its geometry was obtained using both ON(B97-3c:GFN2) and ON($\omega$B97x-3c:GFN2) and then the energy was refined at ON(DLPNO-CCSD(T)/aug-cc-pvtz:GFN2). 
The BE* (uncorrected for the BSSE and ZPE) are reported in Table \ref{tab:bigger_bench}. 
The resulting energies have around 0.3 kJ/mol difference which means that our methodology describes the lower end of the distribution well.
The conclusion of the benchmark was that considering the accuracy and the computational cost, the most optimal methodology for this system is: ON(DLPNO-CCSD(T)/aug-cc-pvtz:GFN2)//ON(B97-3c:GFN2).

\subsection{Zero point energy correction}

\begin{figure}
    \centering
    \includegraphics[width=0.85\linewidth]{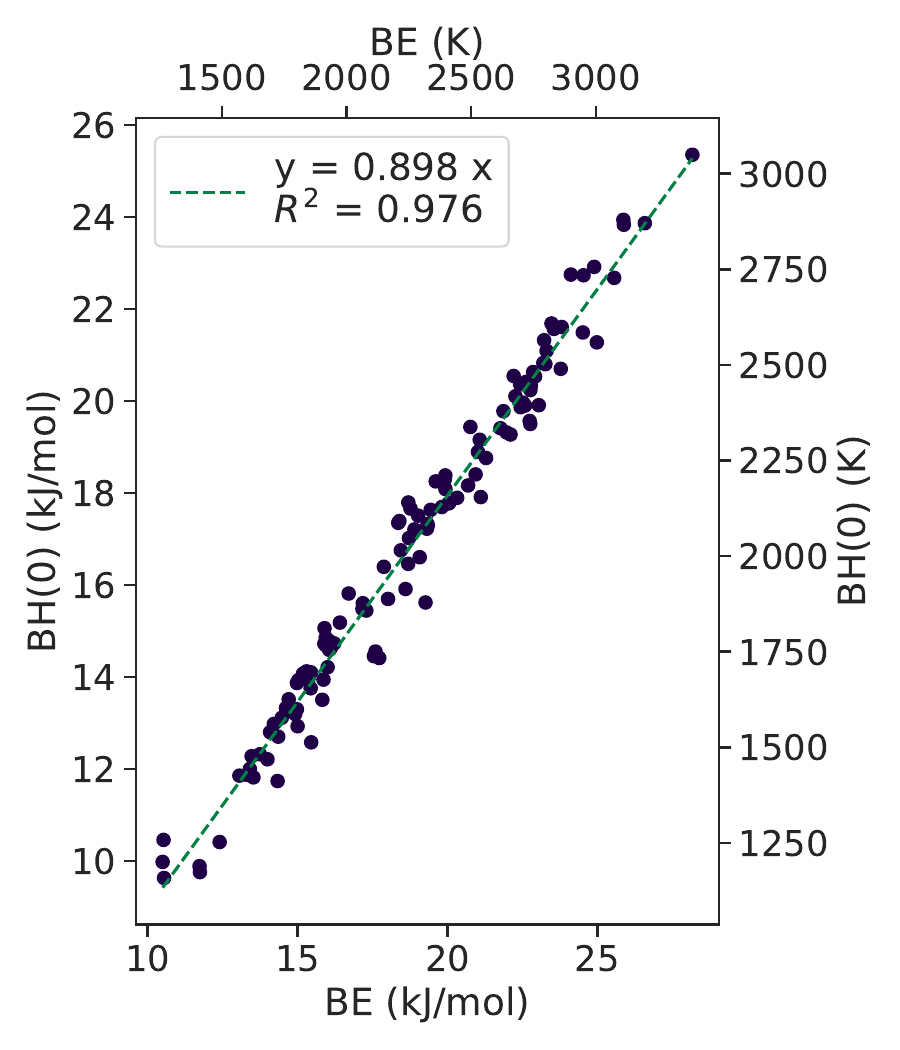}
    \caption{The correlation between the ZPE corrected BE: BH(0) and a non-ZPE corrected BE: BE, that were computed in this study, both in kJ/mol and K along with the linear fit: $\mathrm{BH(0)}=a \cdot \mathrm{BE}$}
    \label{fig:zpe}
\end{figure}

The zero-point energy (ZPE) correction was calculated separately for each structure.
Some previous \citep{Ferrero2020, Perrero2022, Martínez-Bachs_2024, Kakkar_2025} studies used a scaling factor which was derived using the calculations on a crystalline water ice with different types of species.
However, in this approach all species are treated in the same way despite different types of interactions that could affect this property.
To verify this, the correlation between the BE corrected for the ZPE (BH(0)) and the non-corrected BE is plotted in Figure \ref{fig:zpe}.
The scaling factor based on the BEs computed in this work is 0.898, shown in \ref{fig:zpe}, while for a similar system in a previous study done by some of us \citep{Bulik2025} for CO$_2$ on adsorbed on an amorphous water solid the value was 0.754.
This is around 16\% difference and can have a significant impact on the final BE.
This confirms our belief that computing the exact ZPE correction for each binding site is the most reliable approach.

\section{Bootstrap details}

\begin{figure}[h!]
    \centering
    \includegraphics[width=0.95\linewidth]{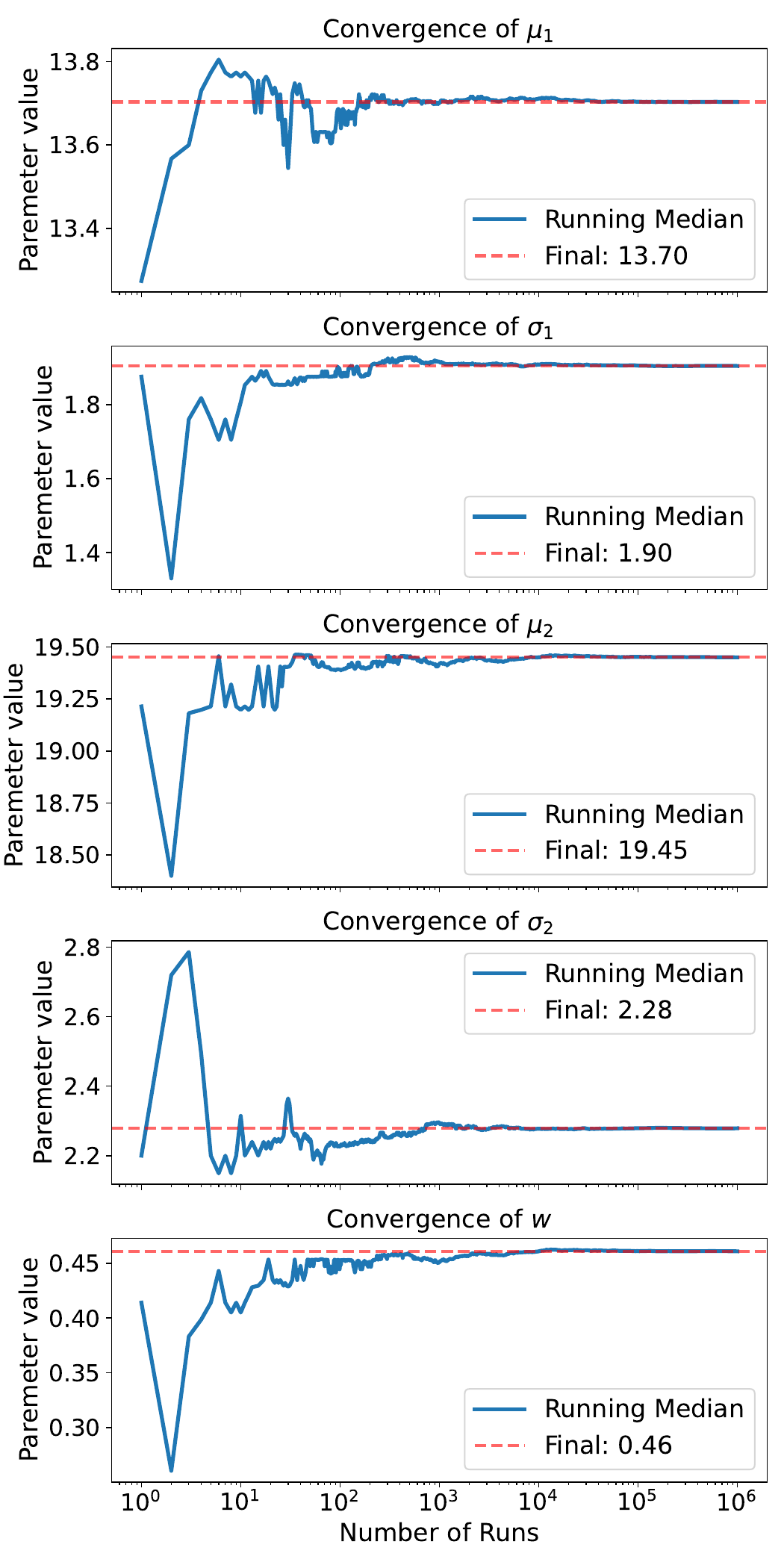}
    \caption{The plot showing the median during of each of the parameters over the course of the bootstrap, the units of the $\mu_i$ and $\sigma_i$ parameters is kJ/mol and the parameter $w$ is dimensionless.}
    \label{fig:bootstrap}
\end{figure}

The bootstrap method was used to estimate the distribution parameters describing the computed BEs.
It was run a 1 million times to ensure the convergence of the parameters.
The parameters themselves were extracted from the parameter estimation procedure by taking the median, with errors defined at 95\% confidence intervals.
The convergence of the bootstrap is shown in Figure \ref{fig:bootstrap}.
The cumulative median was computed at each step of the bootstrap, the red line depicts the final median of each of the parameters.
The plot shows that the 1 million steps was sufficient to get a reliable convergence for each parameter value.
Finally, from the logarithmic ratio test (LRT) we confirmed that the distribution is best described by the a sum of two Gaussian functions, with p<0.05.

\section{Pre-exponential factor formalism}

\begin{figure}[h!]
    \centering
    \includegraphics[width=0.95\linewidth]{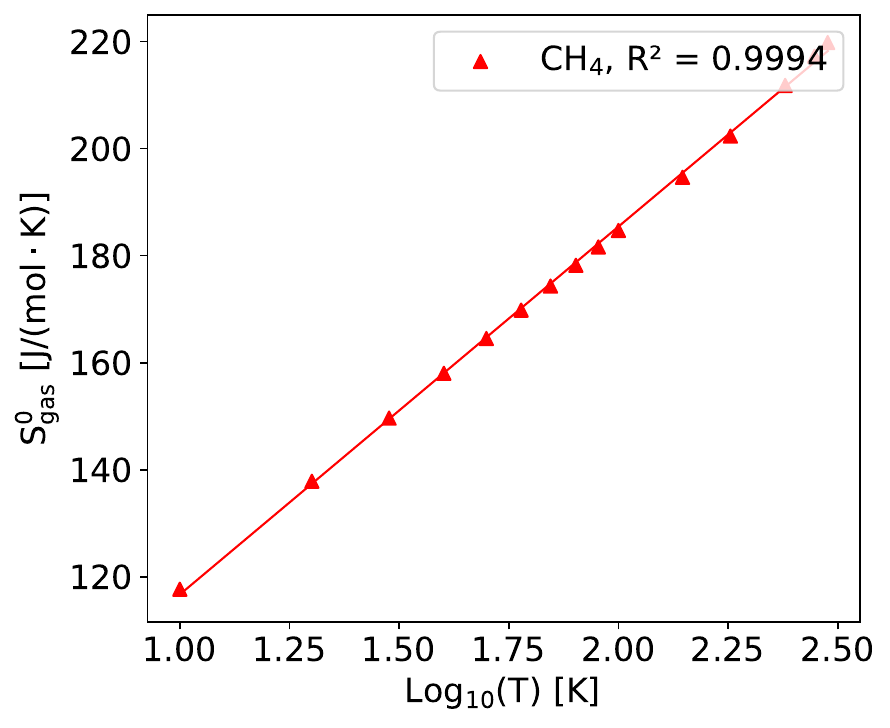}
    \caption{Correlation plot of $S\mathring{}_{\mathrm{gas}}$ and the log$_{10}$T for CO$_2$} 
    \label{fig:entropy_campbell}
\end{figure}

\begin{figure}
    \centering
    \includegraphics[width=0.95\linewidth]{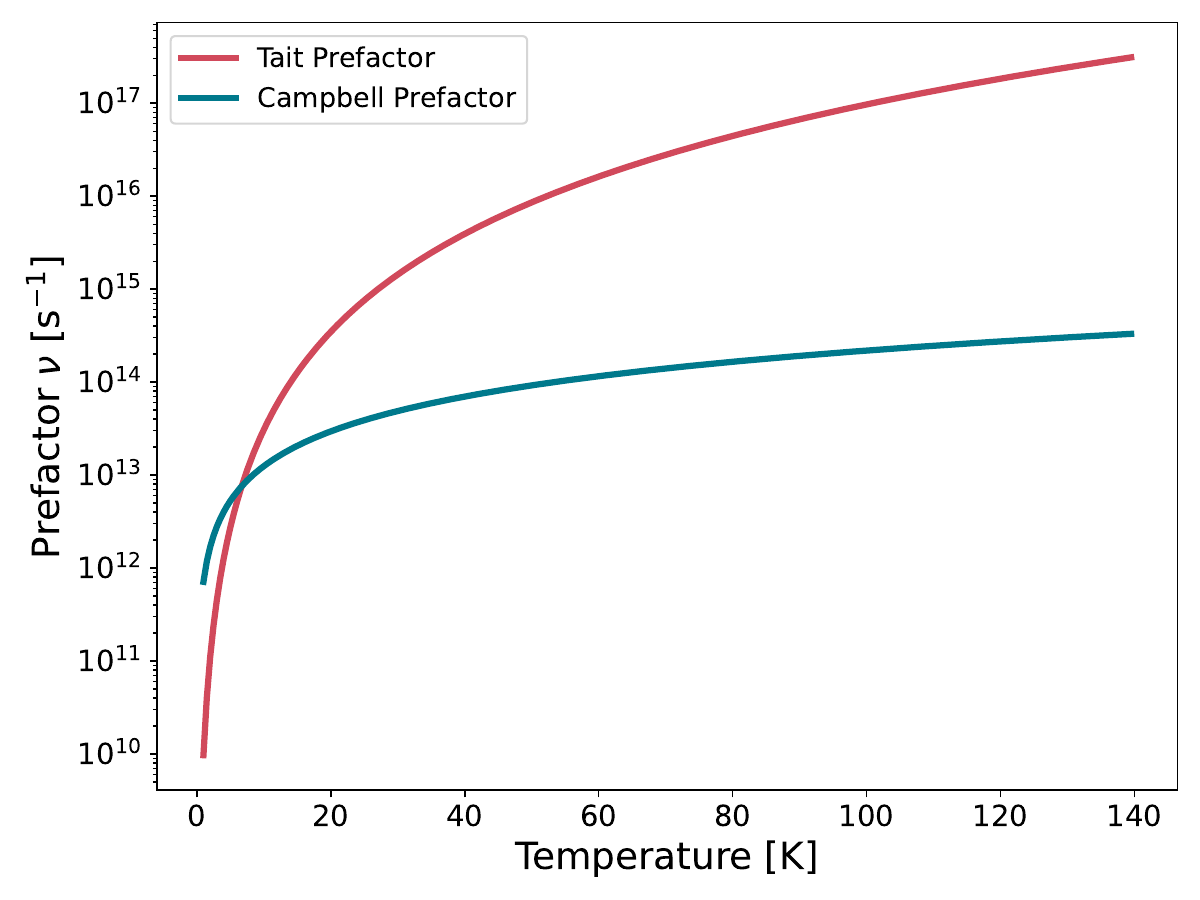}
    \caption{The temperature dependence of each of the prefactor models} 
    \label{fig:pref_temp}
\end{figure}

\begin{table*}[h!]
\caption{Physical constants and parameters used for TPD curve simulation.}
\label{table:constants}
\centering
\begin{tabular*} {\linewidth}{@{\extracolsep{\fill}} l c c c  @{}} 
\hline\hline
Symbol & Description & Value & Units \\ 
\hline
$h$ & Planck constant & $6.62606957\times10^{-34}$ & m$^{2}$\,kg\,s$^{-1}$ \\ 
$k_\mathrm{B}$ & Boltzmann constant & $1.3806488\times10^{-23}$ & m$^{2}$\,kg\,s$^{-2}$\,K$^{-1}$ \\ 
$A$ & Surface per adsorbed molecule & $1\times10^{-19}$ & m$^{2}$ \\ 
$I_x$ & Moment of inertia along $x$  & 1 & m$^{2}$\,kg \\ 
$I_y$ & Moment of inertia along $y$  & 43.448 & m$^{2}$\,kg \\ 
$I_z$ & Moment of inertia along $z$  & 43.448 & m$^{2}$\,kg \\ 
$m_A$ & CO$_2$ mass & $7.35237\times10^{-26}$ & kg \\ 
$m_{\text{AR}}$ & Ar mass & $6.633521\times10^{-26}$ & kg \\
$R$ & Molar gas constant & $8.3144621$ & J\,K$^{-1}$\,mol$^{-1}$ \\ 
\hline
\end{tabular*}
\end{table*}

The BE is extracted from the Polanyi-Wigner ( Polanyi, Wigner 1925) equation.

\begin{equation}
    -\frac{d\theta}{dt} = \theta^{n} \cdot
    \underbrace{v \cdot \exp{\left(-\frac{E_{\mathrm{des}}}{k_B T}\right)}}_{k_{\mathrm{des}}(T)} 
\label{eq:polanyi}
\end{equation}

where $\theta$ represents the coverage, $t$ is time, $n$ is the order of the process, $v$ is the pre-exponential factor (prefactor), $k_B$ is the Boltzmann constant, $T$ is the temperature. 
Finally, $\text{BE}$ is equivalent to $E_{\text{des}}$ and is easily extracted through mathematical operations.  

In this work two prefactor formalisms have been adopted: $v_{Tait}$ \citet{Tait2005,Tait2006} and $v_{Campbell}$ \citet{Campbell2012, Minissale2022}.
The first prefactor, $v_{Tait}$, is described by the following equation:
\begin{equation}
    v_{Tait} = \frac{k_BT}{h} \left(\frac{2\pi m_Ak_BT}{h^2} \right) A \frac{\sqrt{\pi}}{\sigma h^3} (8\pi^2k_BT)^{\frac{3}{2}} \sqrt{I_xI_yI_z}
\end{equation}

The second prefactor: $v_{Campbell}$ is described by the equation:

\begin{equation}
\begin{split}
v_{\mathrm{Campbell}} = \frac{k_B T}{h} 
\exp \Bigg\{
    &\, 0.30\, S\mathring{}_{\mathrm{gas}} + R + 3.3 \\
    &- \frac{1}{3} \Bigg[
        18.6 + \ln \!\left(
            \left( \frac{m_A}{m_{\mathrm{Ar}}} \right)^{3/2}
            \left( \frac{T}{298~\mathrm{K}} \right)^{5/2}
        \right)
    \Bigg]
\Bigg\}
\end{split}
\end{equation}
The constants employed in the calculations along with their units are shown in Table \ref{table:constants}.
For the Campbell prefactor, the $S\mathring{}_{\mathrm{gas}}$ is the entropy of the gaseous species, which can be interpolated with a method reported by \citet{Pantaleone2025}. 
We computed the $S\mathring{}_{\mathrm{gas}}$ values for the species in the gas phase over the temperature range of 10–300K at B97-3c level of theory for the interpolation, see Figure \ref{fig:entropy_campbell}. 
The resulting linear fit is:
\begin{equation}
    S\mathring{}_{\mathrm{gas}} = 68.73 \log_{10}(\text{T}) + 47.96
\end{equation}
Both prefactors are shown as the function of temperature in Figure \ref{fig:pref_temp}. 
\FloatBarrier

\end{appendix}






\end{document}